\documentclass[article,twocolumn,english,footinbib,tightenlines,nobibnotes,aps,unsortedaddress,showpacs,superscriptaddress,secnumarabic]{revtex4-2}
\usepackage{graphicx}
\usepackage[range-units=single,separate-uncertainty]{siunitx}
\usepackage[color links=true,linkcolor=blue,citecolor=blue,urlcolor=blue]{hyperref}

\usepackage[usenames,dvipsnames]{color}
\usepackage{amsmath}
\usepackage{amsfonts}
\usepackage{tikz}
\usepackage{amssymb}
\usepackage{mathtools}
\usepackage{pict2e}
\usepackage{hyperref}
\usepackage{graphicx,xcolor}
\usepackage{bbding}
\usepackage{bm}
\usepackage{bbm}
\usepackage{setspace} %%
\usepackage{multirow}
\usepackage{dcolumn}
\usepackage{float}
\usepackage{amsmath,mathrsfs}
\usepackage{amsthm}
\usepackage{epsfig}
\usepackage{array}
\usepackage{color}
\usepackage[caption=false]{subfig}

\usepackage{comment}

\newcolumntype{P}[1]{>{\centering\arraybackslash}p{#1}}

\def\ie{{\it i.e.\/}}
\def\eg{{\it e.g.\/}}

\mathchardef\sOmega="710A
\mathchardef\sGamma="7100
\mathchardef\sDelta="7101

\newcommand{\supp}{cf.~suppl. }

\begin{document}

% ------------------------------------------------------------------------------
% TITLE AND AUTHORS
% ------------------------------------------------------------------------------

\title{Observation of cnoidal wave localization in non-linear topolectric circuits}

\author{Hendrik Hohmann}
\affiliation{Institute for Theoretical Physics and Astrophysics, University of W\"urzburg, D-97074 W\"urzburg, Germany}
\author{Tobias Hofmann}
\affiliation{Institute for Theoretical Physics and Astrophysics, University of W\"urzburg, D-97074 W\"urzburg, Germany}
\author{Tobias Helbig}
\affiliation{Institute for Theoretical Physics and Astrophysics, University of W\"urzburg, D-97074 W\"urzburg, Germany}
\author{Stefan Imhof}
\affiliation{Physikalisches Institut, Universit\"at W\"urzburg, 97074 W\"urzburg, Germany}
\author{Hauke Brand}
\affiliation{Physikalisches Institut, Universit\"at W\"urzburg, 97074 W\"urzburg, Germany}
\author{Lavi K. Upreti}
\affiliation{Institute for Theoretical Physics and Astrophysics, University of W\"urzburg, D-97074 W\"urzburg, Germany}
\author{Alexander Stegmaier}
\affiliation{Institute for Theoretical Physics and Astrophysics, University of W\"urzburg, D-97074 W\"urzburg, Germany}
\author{Alexander Fritzsche}
\affiliation{Institute for Theoretical Physics and Astrophysics, University of W\"urzburg, D-97074 W\"urzburg, Germany}
\author{Tobias M\"uller}
\affiliation{Institute for Theoretical Physics and Astrophysics, University of W\"urzburg, D-97074 W\"urzburg, Germany}
\author{Udo Schwingenschl\"ogl}
\affiliation{Physical Sciences and Engineering Division (PSE), King Abdullah University of Science and Technology (KAUST), Thuwal 23955-6900, Saudi Arabia}
\author{Ching Hua Lee}
\affiliation{Department of Physics, National University of Singapore, Singapore, 117542}
\author{Martin Greiter}
\affiliation{Institute for Theoretical Physics and Astrophysics, University of W\"urzburg, D-97074 W\"urzburg, Germany}
\author{Laurens W. Molenkamp}
\affiliation{Physikalisches Institut, Universit\"at W\"urzburg, 97074 W\"urzburg, Germany}
\author{Tobias Kie\ss{}ling}
\affiliation{Physikalisches Institut, Universit\"at W\"urzburg, 97074 W\"urzburg, Germany}
\author{Ronny Thomale}
\email{Corresponding author: rthomale@physik.uni-wuerzburg.de}
\affiliation{Institute for Theoretical Physics and Astrophysics, University of W\"urzburg, D-97074 W\"urzburg, Germany}

\date{\today}

% ------------------------------------------------------------------------------
% ABSTRACT
% ------------------------------------------------------------------------------

\begin{abstract}
We observe a localized cnoidal (LCn) state in an electric circuit network. Its formation derives from the interplay of non-linearity and the topology inherent to a Su-Schrieffer-Heeger (SSH) chain of inductors.
Varicap diodes act as voltage-dependent capacitors, and create a non-linear on-site potential. For a sinusoidal voltage excitation around midgap frequency, we show that the voltage response in the non-linear SSH circuit follows the Korteweg-de Vries equation. The topological SSH boundary state which relates to a midgap impedance peak in the linearized limit is distorted into the LCn state in the non-linear regime, where the cnoidal eccentricity decreases from edge to bulk.\end{abstract}
%-------------------------------------------------------------------------------

\maketitle

% ------------------------------------------------------------------------------
%  INTRODUCTION
% ------------------------------------------------------------------------------

\paragraph*{Introduction.}

Since the first observation of a solitary wave in a canal near Edinburgh by J. S. Russel in 1834, solitons as special solutions to non-linear equations of motion have been studied extensively~\cite{Russell,doi:10.1080/14786449508620739,1451226}.
The phenomenon has subsequently gained relevance for a multitude of mathematical, biological, and physical domains~\cite{doi:10.1137/1.9781611970883,PhysRevA.27.2120,Heimburg9790,inbook}.
It includes, but is not exhausted by, hydrodynamic waves in oceans, rivers, and the atmosphere~\cite{doi:10.1126/science.208.4443.451, 10.2307/24966708, EquatorialSolitaryWavesPartIRossbySolitons},
ion-acoustic solitons in plasma~\cite{Lonngren_1983}, and DNA fluctuations~\cite{PhysRevA.27.2120}.
The ability to analytically retrace the soliton has stimulated new developments in optical fiber communications~\cite{doi:10.1063/1.1654836,EMPLIT1987374,RevModPhys.68.423} and solid state physics~\cite{doi:10.1126/science.287.5450.97}, which culminated in the modelling of electrical conductance in polymers through the Su-Schrieffer-Heeger (SSH)
model~\cite{PhysRevLett.42.1698}.
There, the conductivity of trans-polyacetylene derives from charged solitons, propagating as domain walls between two allowed energy configurations~\cite{PhysRevB.22.2099,RevModPhys.60.781}.
Initially investigated in the non-linear soliton regime, a linearized tight-binding description of the SSH model subsequently gained importance as a toy model and building block for symmetry-protected boundary modes and topological phases. In some aspects, the SSH model can be thought of as the cradle of topological classifications, which has substantially deepened the understanding of topological states of matter~\cite{RevModPhys.83.1057,RevModPhys.82.3045}.

Many of the topological phases known to date are not reserved to quantum systems, but rather have additionally or exclusively been realized in classical metamaterials~\cite{PhysRevLett.100.013904}, such as mechanic~\cite{doi:10.1126/science.aab0239}, acoustic~\cite{articlec,TopologicalSound},
photonic~\cite{RevModPhys.91.015006}, and electric~\cite{PhysRevLett.114.173902,PhysRevX.5.021031} setups.
Furthermore, metamaterials have been employed to investigate effects caused by non-linearity, which is
either inherent to the platform, \eg~Kerr non-linearity in optical materials~\cite{PhysRevLett.81.3383}, or added to the setup by non-linear components~\cite{article,FirstKdVCircuit}.
Here, topolectric circuits~\cite{Lee_2018} stand out in terms of versatility and accessibility, as they are capable of realizing both discrete (lattice type) and continuous (transmission line) systems in the short and long wavelength limit, respectively.
Both regimes have been explored experimentally, from topological phases~\cite{Imhof_2018,Hofmann_2019} and band structures~\cite{Helbig_2019,PhysRevB.99.020302} to
solitons~\cite{article,FirstKdVCircuit,Kofan1988,PhysRevE.49.828} and cnoidal waves as periodic soliton-like
solutions of the Korteweg-de Vries (KdV) equation~\cite{doi:10.1080/14786449508620739,Toda,articleee,article}. Moreover, the availability of commercially refined non-linear electronic components, such as non-linear resistors~\cite{Kotwale2106411118} or varicap diodes~\cite{article}, renders electric circuits ideally suited to
study the interplay of topology and non-linearity~\mbox{\cite{Alu1,articlea,doi:10.1021/acsphotonics.7b00303,ChingHua,ezawa2021topological}. }

In this Letter, we investigate a non-linear SSH circuit. In contrast to previous works, where the non-linearity enters in the kinetic term~\cite{Alu1,articlea,doi:10.1021/acsphotonics.7b00303,ChingHua}, we introduce it as a local potential. As a response to a sinusoidal voltage excitation at midgap frequencies fed into the edge of the circuit, we measure the localized cnoidal (LCn) state. Originating from the intertwining of topological localization and non-linearity, the LCn state shows an exponential decay of its root mean square (RMS) amplitude towards the bulk, whereas the non-linearity manifests itself in a temporal distortion of the sinusoidal character of the input.
We develop an approach to theoretically describe the LCn state by separating the chain into decoupled LC-resonators which are described by the KdV equation and its cnoidal wave solutions.
Our findings establish circuit networks as the platform of choice to explore non-linear topological matter.
%-------------------------------------------------------------------------------

% ------------------------------------------------------------------------------
%  MAIN
% ------------------------------------------------------------------------------

%-------------------------------------------------------------------------------
% Figure 1
%-------------------------------------------------------------------------------
\begin{figure*}
    \centering
    \begin{tikzpicture}
        \node[inner sep=0pt] at (2.5, 1.3)
{\includegraphics[width=0.89\textwidth]{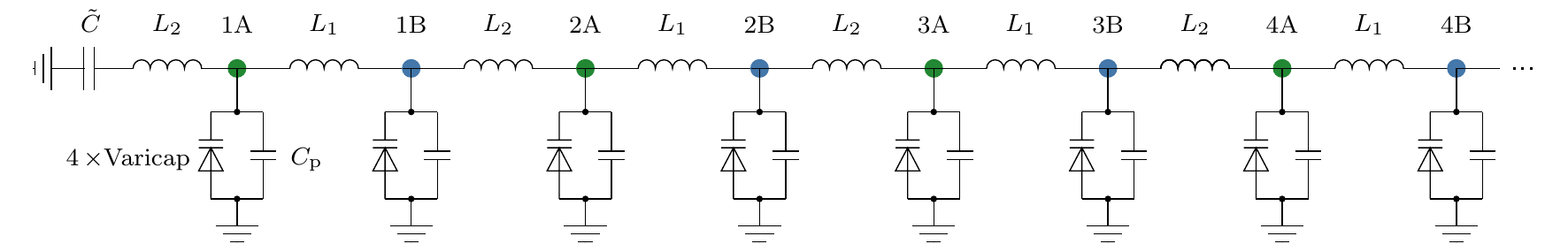}};

        \node[inner sep=0pt] at (-3.3,  -2.375)
{\includegraphics[width=0.246\textwidth]{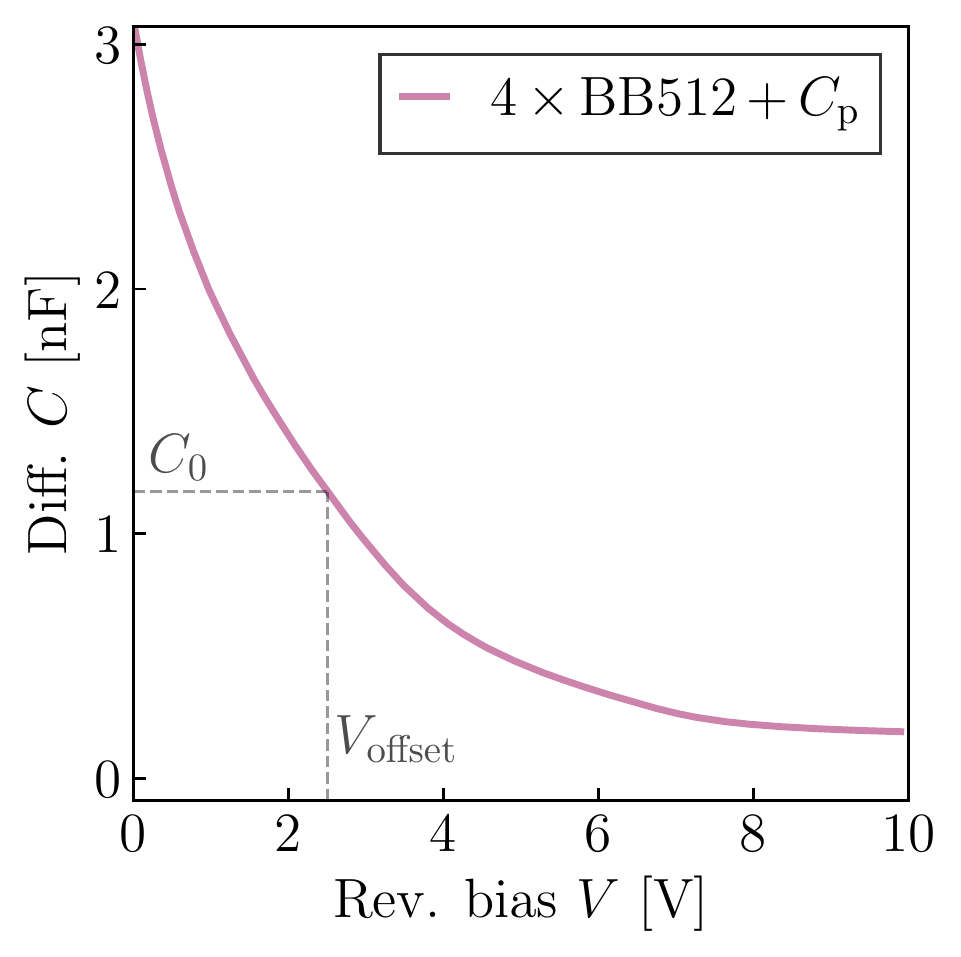}};
        \node[inner sep=0pt] at (7, -2.45)
{\includegraphics[width=0.367\textwidth]{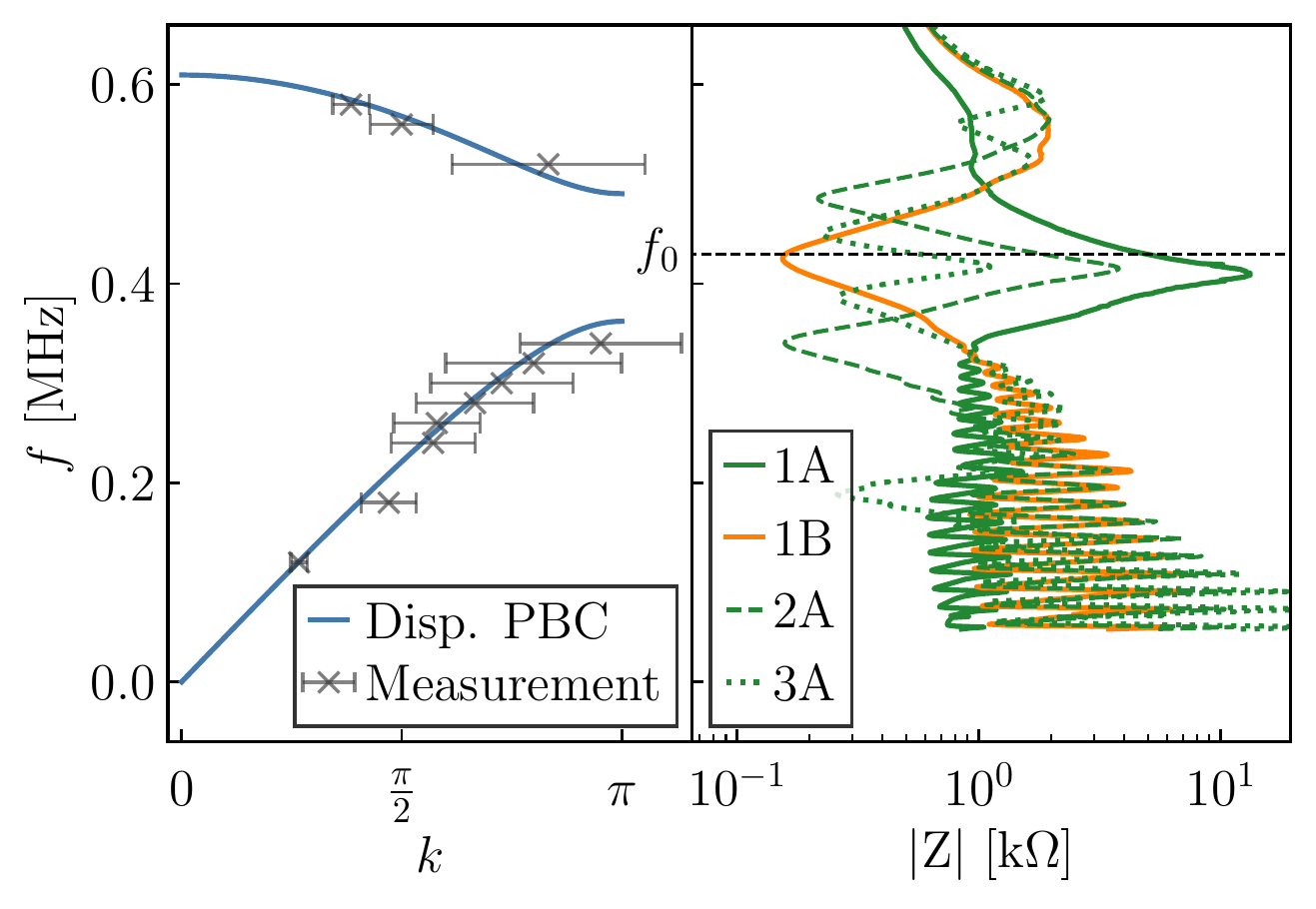}};
        \node[inner sep=0pt] at (1.4, -2.08)
{\includegraphics[width=0.22\textwidth]{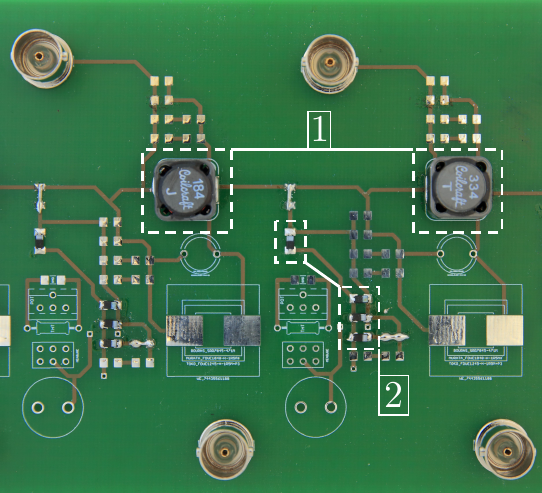}};

\draw[rounded corners] (-4.9,0.4) rectangle (-2.1,1.5);

\draw[gray, rounded corners] (0.03,-0.05) rectangle (3.57,2.55);

\node at (1.35, 0.12)  { \footnotesize{unit cell} };

        \node at (-5.35, 2.4)  { (a) };
        \node at (-5.35, -0.5)  { (b) };
        \node at (-0.87, -0.5)  { (c) };
        \node at (3.85, -0.5)  { (d) };
        \node at (-4.25, 0.23)  { \footnotesize{non-linear} };
    \end{tikzpicture}
    \caption{(a)~Schematic of the nSSH circuit. Alternating inductances $L_1$ and $L_2$ connect the voltage nodes, reverse-biased varicap diodes to ground act as non-linear capacitances and the parallel capacitor $C_{\text{p}}$ takes the parasitic influence of the measurement setup into account (\supp~D). $\widetilde{C} \gg C(V_{\text{offset}})$ blocks the DC voltage offset. The gray rectangle depicts one unit cell. (b)~Measurement of the voltage dependence of the four Siemens BB512 varicap diodes in parallel configuration and $C_{\text{p}}$. (c)~One unit cell of the nSSH circuit on the PCB with [1] inductors of nominal values $L_{1,\text{nom}} = \SI{330}{\micro\henry}$ and $L_{2,\text{nom}} = \SI{180}{\micro\henry}$, and [2] varicap diodes.
		(d)~(left) Theoretical (blue curve) and measured (grey crosses) dispersion relation for PBC in the linear limit at the operating point $V_{\text{offset}} = \SI{2.5}{\volt}$. (right) Small signal impedance analysis between nodes 1A, 1B, 2A, 3A and ground for OBC. An impedance peak at the midgap frequency $f_0 = \SI{430}{\kilo\hertz}$ indicates the localized SSH boundary state. } 
		    \label{fig:circuit}
\end{figure*}
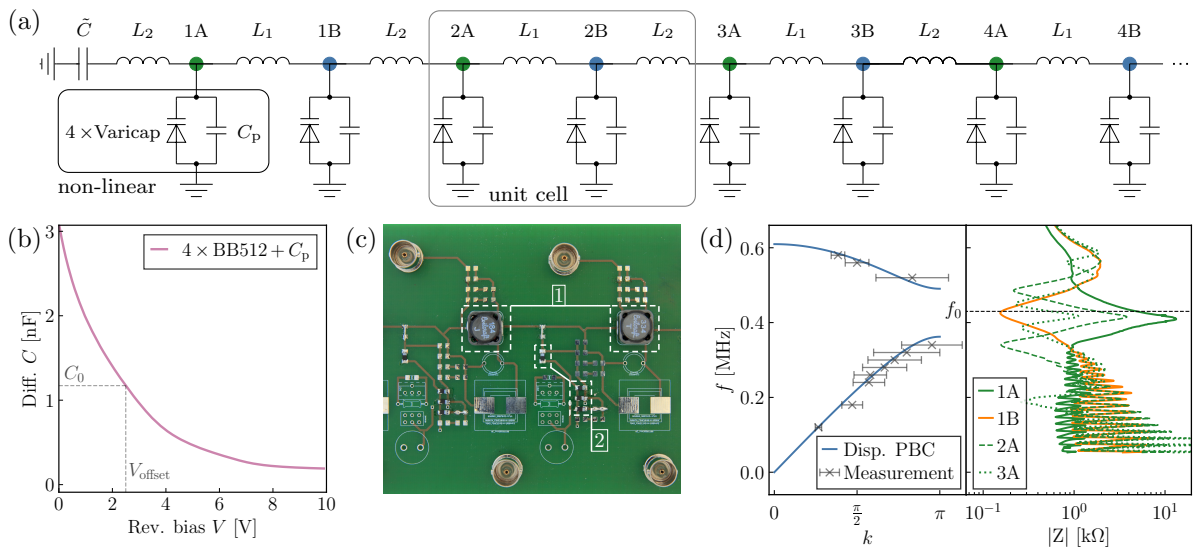
%-------------------------------------------------------------------------------

%-------------------------------------------------------------------------------
% Paragraph: Circuit model
%-------------------------------------------------------------------------------

\paragraph*{Non-linear SSH circuit.}
We create a non-linear SSH (nSSH) circuit, schematically depicted in Fig.~\ref{fig:circuit}~(a), in which alternating inductances $L_1$ and $L_2$ connect the circuit nodes. Each node in the two-site unit cell is grounded by a parallel configuration of four varicap diodes of type Siemens BB512 in reverse-bias setting and a linear capacitor $C_\text{p}$. They realize a non-linear on-site capacitance and extend the linear SSH model to the non-linear regime. $C_\text{p}$ is added to the theoretical model to account for parasitic capacitances induced by the measurement setup. From the actual printed circuit board (PCB) with a total of 25 unit cells and 50 voltage nodes, a cutout of one unit cell is shown in Fig.~\ref{fig:circuit}~(c).
Since the varicap diodes would be conductive for negative node voltages, we operate the circuit at positive voltages in the depletion region. This is achieved by applying a DC voltage offset $V_{\text{offset}}$ to all nodes, which defines the operating point of the non-linear capacitance $C_0 = C(V_{\text{offset}})$. In order to experimentally stabilize the voltage offset, the nodes at the edges are decoupled from ground by a large additional capacitor $\widetilde{C} \gg C_0$, which does not affect the dynamical behavior of the system. The total differential capacitance $C(V)$ from each node to ground decreases non-linearly as a function of the reverse biased voltage $V$, measured in Fig.~\ref{fig:circuit}~(b).
In addition to the voltage offset, we excite the nSSH circuit with an AC voltage signal yielding a total input voltage $V(t) = V_{\text{offset}} + A_0\,\sin(\omega t)$. In this passive circuit, the effect of the non-linearity on the voltages and currents increases with larger excitation amplitudes $A_0$. This allows us to employ the amplitude of the AC input signal as a tuning parameter for the influence of the non-linear on-site capacitance on the circuit.
%-------------------------------------------------------------------------------

%-------------------------------------------------------------------------------
% Paragraph: Linear limit
%-------------------------------------------------------------------------------

\paragraph*{Linear limit of the nSSH circuit.}
To connect the spatial character of the LCn state to the description of the nSHH circuit in its linear limit, we employ a small amplitude signal analysis. For small AC input signals ($A_0 \ll V_{\text{offset}}$), we assume the on-site capacitance to be constant, $C(V) \approx C_0$, and linearize the equations of motion (EOM) around the operating point $V_{\text{offset}}$ to obtain an effective description in the linear limit. The linearized EOM are diagonal in frequency space and their solutions are characterized by the dispersion relation.

The effective linear model is equivalent to the SSH chain, and the dispersion relation $\omega(k)$ for periodic boundary conditions (PBC) is given by
\begin{align}\label{eq:dispersion}
	\omega_{\pm}^2(k) = \omega_0^2 \pm \frac{1}{C_0} \sqrt{\frac{1}{L_1^2} + \frac{1}{L_2^2} + \frac{2}{L_1\,L_2}\,\cos(k)} ,
\end{align}
with $\omega_0^2  = C_0^{-1} \, (L_1^{-1} + L_2^{-1})\equiv (2 \pi f_0)^2$. We set $L_1 > L_2 $ to tune the circuit into the topological regime.

We measure the dispersion relation $\omega(k)$ in the nSSH circuit by reading out one wavelength of the real space voltage distribution resulting from a small signal excitation with frequency $\omega$ (c.f. left part of Fig.~\ref{fig:circuit}~(d)). Due to the finite resolution in real space, the precision decreases for $k \rightarrow \pi$.
In the right part of Fig.~\ref{fig:circuit}~(d) we depict a frequency-resolved small-signal impedance measurement between nodes close to the boundary and ground for open boundary conditions (OBC). Within the two branches of the dispersion relation, we identify several impedance peaks corresponding to the individual bulk modes of the finite circuit network.
Our measurements reveal the boundary state of the SSH model at the mid-gap frequency $f_0$, featuring localization on sublattice A at the edge and exponential decay towards the bulk. 
%-------------------------------------------------------------------------------

%-------------------------------------------------------------------------------
% Figure 2
%-------------------------------------------------------------------------------
\begin{figure}
    \centering
    \begin{tikzpicture}
        \node[inner sep=0pt] at (-4.2, 0)
{\includegraphics[width=0.49\textwidth]{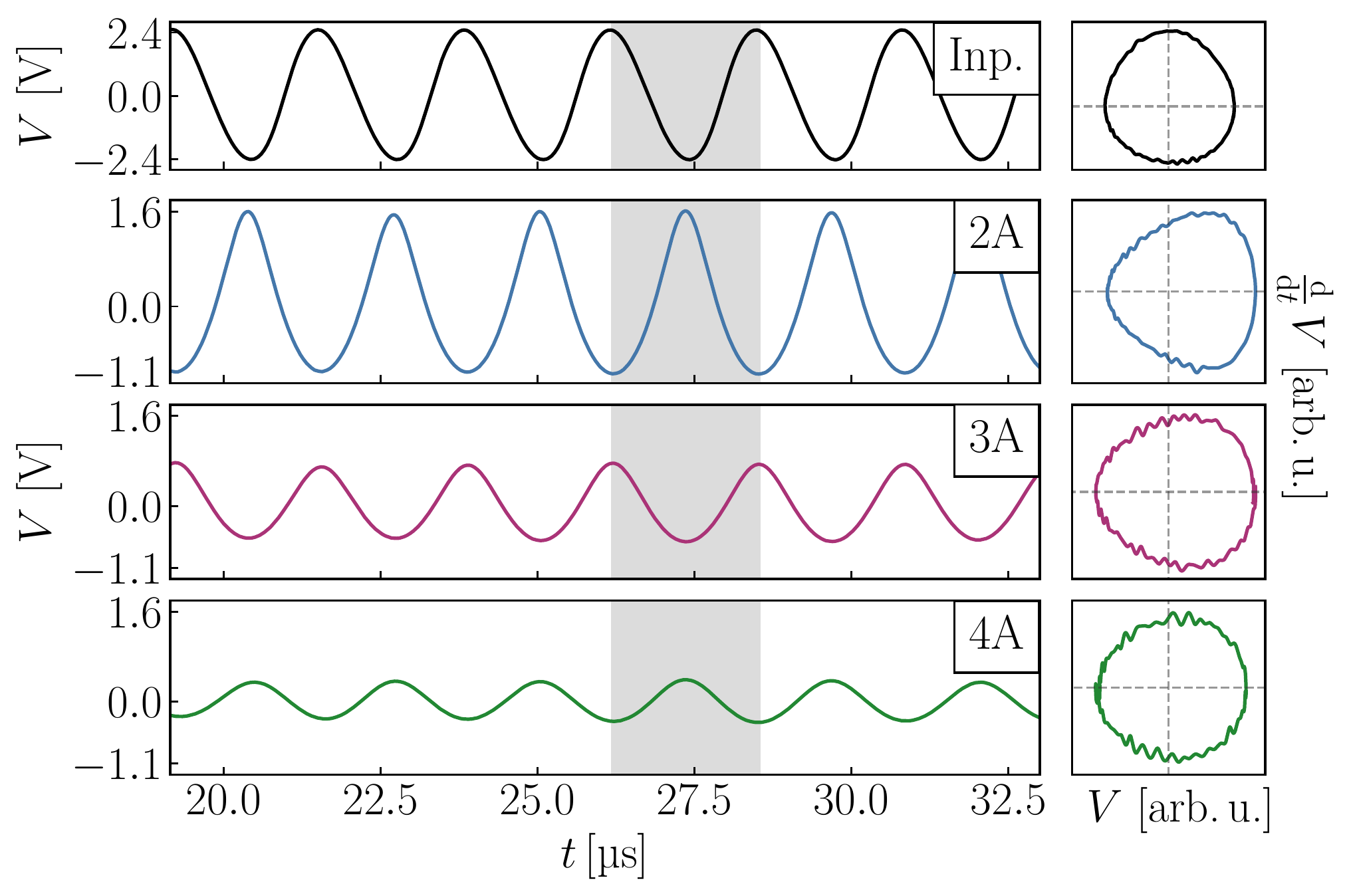}};
    \end{tikzpicture}
    \caption{AC measurement of the LCn state at $V_\text{offset}=\SI{2.5}{\volt}$.
		(top)~Sinusoidal input signal with frequency $f_0 = \SI{430}{\kilo\hertz}$ and amplitude $A_0=\SI{2.5}{\volt}$ fed into node 1A.
		(left)~Steady state voltage response at nodes 2A, 3A, and 4A starting at $t \approx \SI{18}{\micro\second}$. The sinusoidal input is deformed into the LCn state, where amplitude and eccentricity decrease from edge to bulk.
		(right)~Phase space plot of one period of the voltage signal (indicated by the gray area in (a)), visualizing its deformation.
		}
    \label{fig:measurement}
\end{figure}
%-------------------------------------------------------------------------------

%-------------------------------------------------------------------------------
% Paragraph: Continuum limit
%-------------------------------------------------------------------------------

\paragraph*{Continuum limit: Non-linear transmission line.}

The complete non-linear circuit EOM at large input amplitudes $A_0$ on the order of $V_{\text{offset}}$ can not be solved exactly. A standard approach is to apply a continuum approximation by placing oneself in the lower branch of the dispersion in Fig.~\ref{fig:circuit}~(d) close to $k = 0$ with no phase shift between sublattice sites. In this low energy, long-wavelength limit, the dimerization and discrete lattice character are no longer relevant and the circuit setup realizes a non-linear transmission line~\cite{Kofan1988}.
The EOM resemble the KdV equation (\supp~A), for which a bell-shaped excitation leads to the formation of a pulse soliton~\cite{FirstKdVCircuit}. It propagates with constant shape and velocity through the transmission line, because the defocusing effect of the dispersion relation is compensated by the non-linearity.
This approach fails for regions close to the band gap. The phase shift of $\pi$ between adjacent unit cells as well as the suppressed signal on sublattice B in the topological mid-gap state invalidate the continuum approximation.

%-------------------------------------------------------------------------------

%-------------------------------------------------------------------------------
% Figure 3
%-------------------------------------------------------------------------------
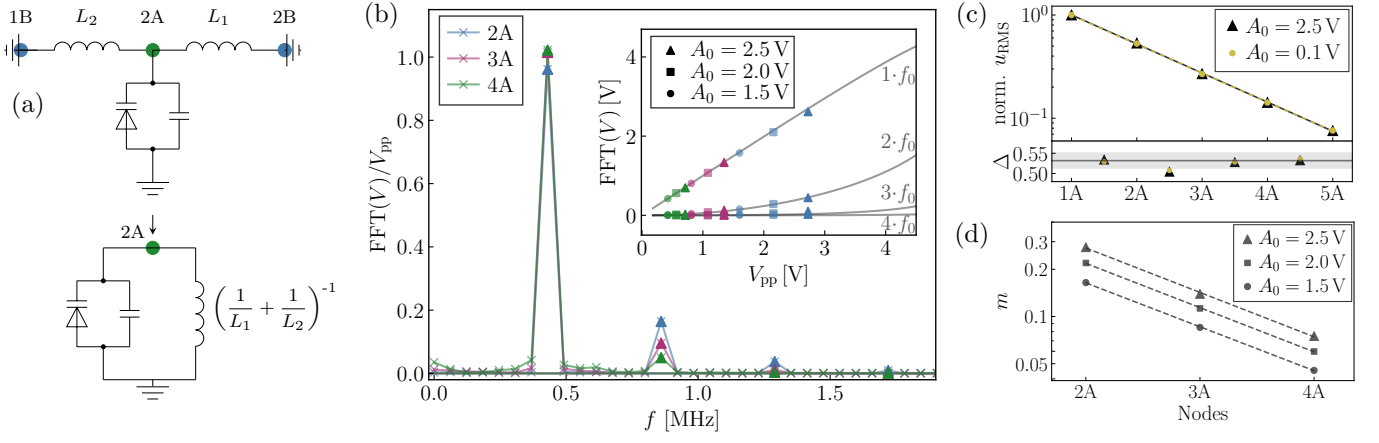
\begin{figure*}
    \centering
    \begin{tikzpicture}
        \node[inner sep=0pt] at (-10, 0.25)
    {\includegraphics[width=0.25\textwidth]{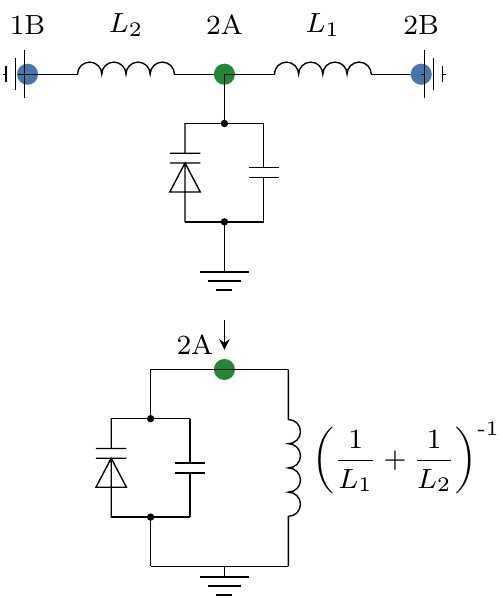}};
        \node[inner sep=0pt] at (-3.6, 0)
{\includegraphics[width=0.435\textwidth]{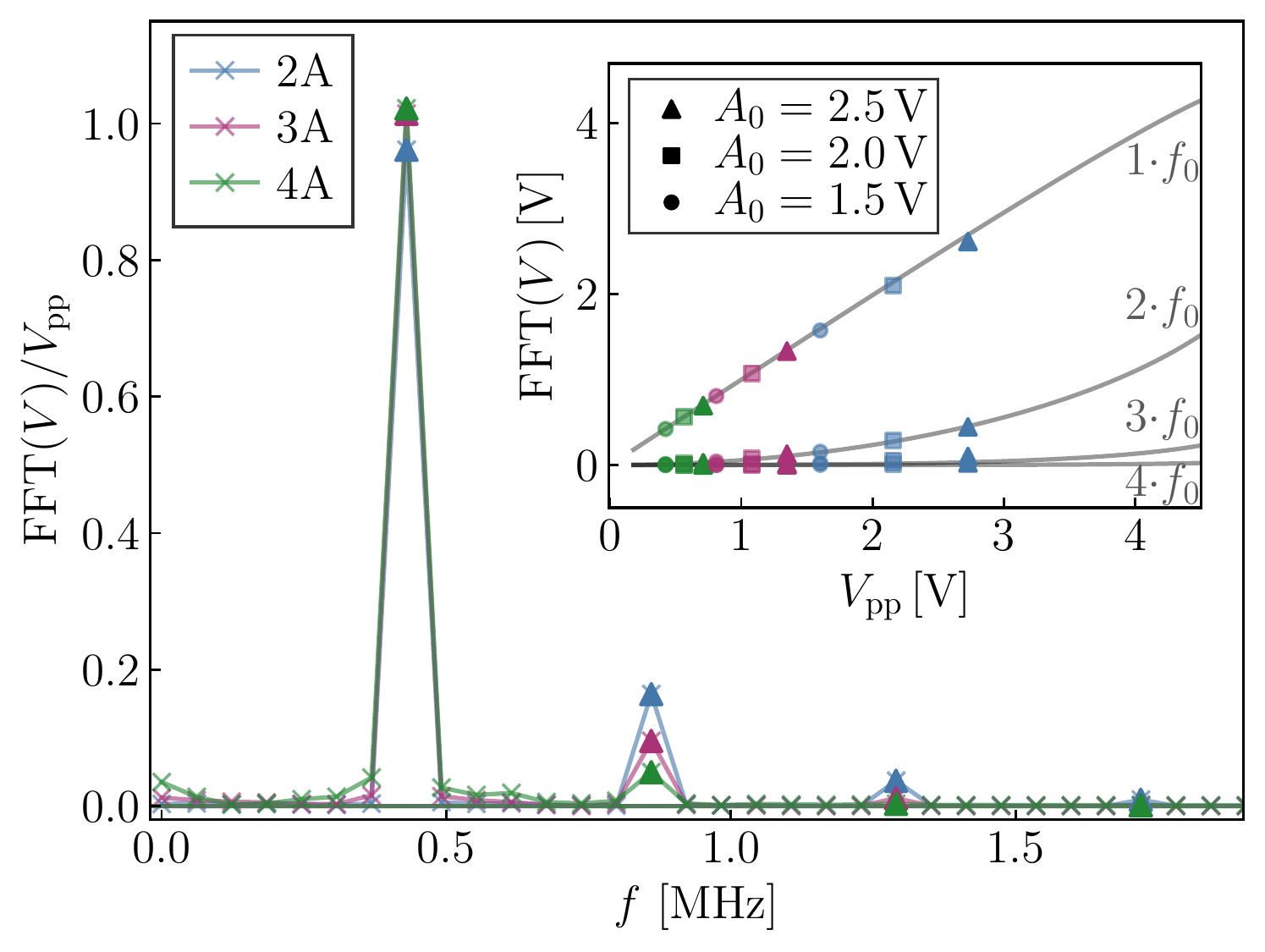}};
        \node[inner sep=0pt] at (3.328, -1.37)
{\includegraphics[width=0.279\textwidth]{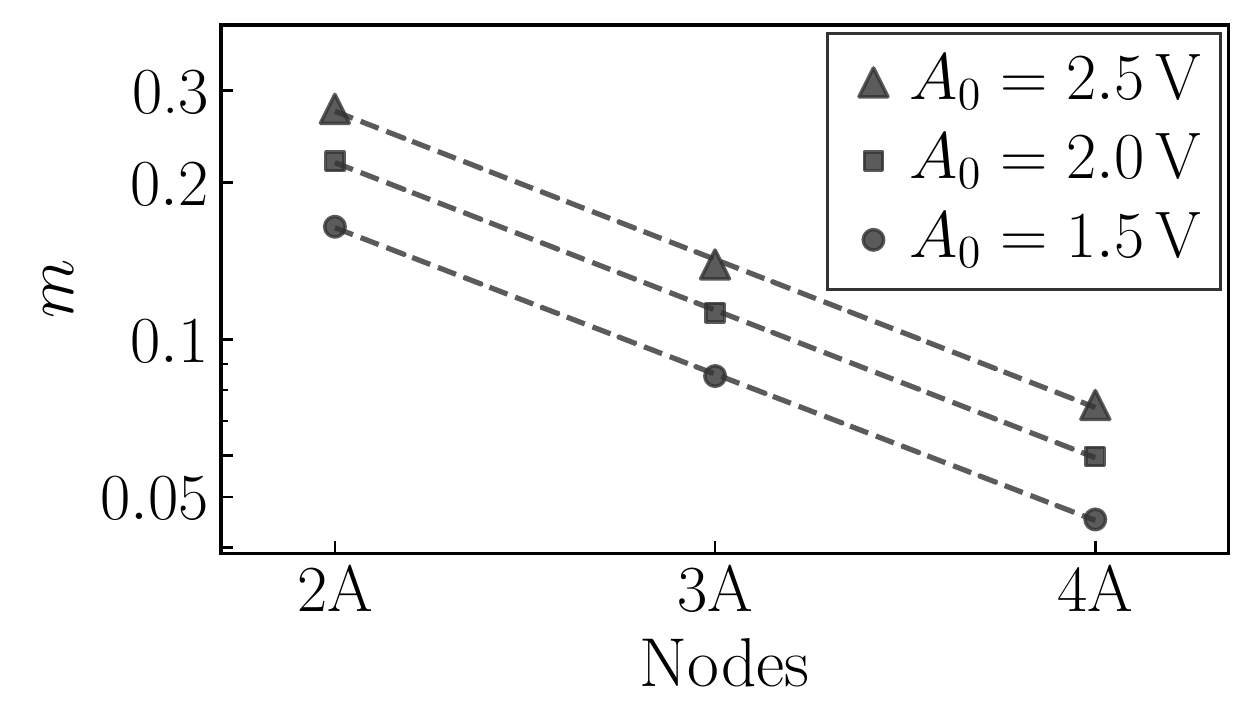}};
       \node[inner sep=0pt] at (3.3, 1.53)
{\includegraphics[width=0.28\textwidth]{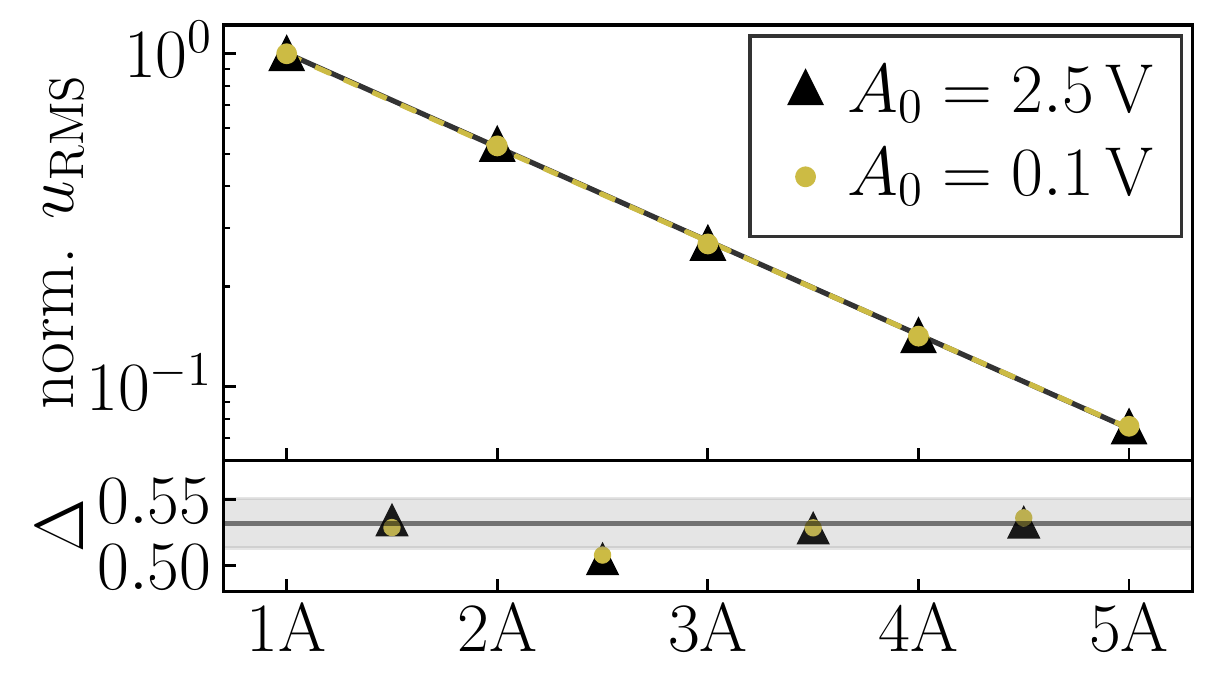}};
	\node at (-11.9, 1.53)  { (a) };
        \node at (-7.2, 2.73)  { (b) };
        \node at (0.65, 2.73)  { (c) };
         \node at (0.65, -0.2)  { (d) };
    \end{tikzpicture}
 \caption{(a)~Transformation of a segment of the nSSH circuit to a non-linear LC resonator by treating the sublattice nodes B as virtual ground, due to their negligible amplitude in the LCn state. (b)~Discrete Fourier transformation of the measured voltage response in the LCn state at first three sublattice A nodes (color encoded), normalized w.r.t. the peak to peak voltage $V_{\text{pp}}$. The crosses mark the evaluated values, guided by solid lines and with triangles at multiples of the base frequency $f_0$. Due to non-linear effects, higher harmonics of the fundamental frequency $f_0 = \SI{430}{\kilo\hertz}$ are excited. The inset compares the Fourier components for different excitation amplitudes $A_0$ (marker encoded), and nodes 2A, 3A, and 4A, with the analytical solution in Eq.~\ref{eq:single_resonator_solution}, depicted as grey lines. (c)~Logarithmic plot of the normalized RMS voltage for the linear and non-linear case with respective linear fits. The ratio $\Delta$ between subsequent RMS values together with the theoretically expected value in grey is shown below. (d)~Logarithmic plot of the eccentricity $m$ for different excitation amplitudes $A_0$ with linear fits.
}
    \label{fig:evaluation}
\end{figure*}
%-------------------------------------------------------------------------------

%-------------------------------------------------------------------------------
% Paragraph: Localized cnoidal state START
%-------------------------------------------------------------------------------

\paragraph*{Dimerized nSSH circuit: Localized cnoidal state.}
We excite the nSSH circuit with a sinusoidal signal of mid-gap frequency $f_0$ at the boundary and measure the voltage response once a steady state is reached. With the localization on sublattice A in the linear topological edge state, non-linear effects are strongly suppressed at sublattice B, resulting in their on-site amplitudes to remain small even for large $A_0$ (\supp B).
Figure~\ref{fig:measurement} shows the voltage response of sublattice nodes A to a sinusoidal excitation at node 1A in the non-linear regime ($A_0 = \SI{2.5}{\volt}$).
We identify this voltage configuration with spatial localization at the boundary in the non-linear regime as the LCn state.
The effect of non-linearity manifests itself in the temporal distortion of the sinusoidal signal at nodes of sublattice A: The phase space plots on the right side of Fig.~\ref{fig:measurement} show how the voltage response is deformed as compared to the elliptic shape of the input signal. The eccentricity, a parameter characterizing the signal distortion with respect to a sinusoidal wave, decreases towards the bulk.

As the frequency $f_0$ of the driving sets the period of the voltage response in the whole circuit, each waveform is composed of higher harmonics at integer multiples of this base frequency. Due to the band gap around $f_0$ and since there are no dispersive states at higher harmonics of $f_0$, it is the LCn mode that is excited predominantly.
%-------------------------------------------------------------------------------

%-------------------------------------------------------------------------------
% Paragraph: Localized cnoidal state CONT'D
%-------------------------------------------------------------------------------

\paragraph*{Theoretical description of the LCn state.}
Given the small AC amplitude of sublattice nodes B in the midgap state, we can approximate them as being replaced by AC ground, as shown in Fig.~\ref{fig:evaluation}~(a). This decouples the chain into a set of independent single resonators, reducing the full non-linear differential equations to a local homogeneous subset (\supp~B, C).
The amplitude remains as an undetermined parameter in the solution of the reduced EOM and is specified by the spatial voltage profile of the LCn state.
Exemplified in Fig.~\ref{fig:evaluation}~(a) for node 2A, each single resonator can be recast as an LC circuit with inductance $L = (L_1^{-1} + L_2^{-1})^{-1}$ and non-linear capacitance $C(V)$.

In the linear limit, the resonance frequency of the single resonator matches that of the midgap SSH edge state with $f_0 = 1/(2 \pi \sqrt{L C_0} )$. This originates from the dimerized regime of the tight-binding model, where the topological state resides at zero energy within the band gap. The analogous scale of zero energy in electric circuits is given by the midgap frequency $f_0$.

To analytically trace the non-linear differential equations of the single resonator, we model the voltage-dependent differential capacitance at voltages $V$ close to $V_{\text{offset}}$ as an inverted square root law \cite{PhysRevLett.47.1349}
\begin{align}\label{eq:capacitance_model}
	C(V) \approx \frac{\mathcal{C}}{\sqrt{1 + \frac{V-\nu}{\phi}}} ,
\end{align}
where $\phi$, $\mathcal{C}$, and $\nu$ are fit parameters to the measured capacitance in Fig.~\ref{fig:circuit}~(b) (\supp~C).
Equation \eqref{eq:capacitance_model} is based on the relation for the differential capacitance of junction diodes.

Since we are only interested in its AC contribution, we decompose the voltage as $V(t) = V_{\text{offset}} + u(t)$.
The non-linear relation $Q(u(t))$ between the charge $Q(t)$ accumulated on the capacitor and the voltage $u(t)$ across it is obtained by integrating Eq.~\ref{eq:capacitance_model} using $\text{d}Q = C(V) \text{d}V$. Inserting the inverted function $u(Q(t))$ into the standard differential equation for the LC resonator, we find that the charge $Q(t)$ follows the KdV equation in its stationary limit~\cite{phdthesis}. The periodic solutions for $Q(t)$ are given by cnoidal waves~\cite{BLYUSS200229} which can be regarded as a periodic arrangement of single solitons.
With the relation $u(Q(t))$, we obtain the analytic description of the AC voltage
\begin{align}\label{eq:single_resonator_solution}
u(t) = \dfrac{1}{4\, \mathcal{C}^2 \,\phi} \, \big[ & \eta \, (\eta + 2 \Theta)  + 2 A_Q\, (\eta + \Theta) \, \text{cn}^2\bigr(\mu t  \, |\, m\bigr) \nonumber  \\
&+ A_Q^2 \, \text{cn}^4\bigr(\mu t \, | \, m\bigr)  \big],
\end{align}
where $\text{cn}(x \,|\, m)$ denotes the Jacobi elliptic cosine function, $m \in [0,1]$ the eccentricity of the wave, and $\Theta$ a system-dependent constant. For the functional dependencies of the valley elevation $\eta(m)$, the peak to peak amplitude $A_{Q}(m)$ of the cnoidal wave solution $Q(t)$, the elliptic frequency $\mu(m)$ on the parameter $m$, and a full derivation refer to supplemental material~C.
The period of $\text{cn}^2\bigr(\mu t  \, |\, m\bigr)$ is $T = 2 K(m)/\mu(m)$ and fixed by the external input frequency $f_0$. $K(m)$ denotes the complete elliptic integral of the first kind.
For eccentricities up to $m\approx 0.3$, the eigenfrequency of the single resonator $f(m)=1/T(m)$ is approximately constant and matches the excitation frequency (\supp~C). Hence, the solution Eq. \eqref{eq:single_resonator_solution} applies to the driven setup.

In Fig.~\ref{fig:evaluation}~(b) we perform a discrete Fourier transformation of the measured steady state for an input amplitude $A_0$ of $2.5\,$V. Due to the non-linearity, higher harmonics of the fundamental frequency $f_0$ are excited, albeit with smaller amplitude. In agreement with the time scale induced by the driving, the EOM for the single resonator allow for solutions which are composed of excitations at multiples of the base frequency $f_0$. As shown in the inset of Fig.~\ref{fig:evaluation}~(b), the measured data agrees with the Fourier coefficients of the theoretical AC voltage solution in
Eq.~\ref{eq:single_resonator_solution} depicted in grey.
The spatially resolved normalized RMS value $u_\text{RMS}$ of the LCn state is shown in Fig.~\ref{fig:evaluation}~(c) and compared to the linear limit. It decreases exponentially towards the bulk with $u_\text{RMS} \propto \Delta^x$ with the same attenuation factor $\Delta$ in the linear and non-linear regime, where $x$ denotes the number of unit cells counted from the edge.
The mean experimental values are $\Delta_\text{lin} = 0.5237(17) $ and $\Delta_\text{n-lin} = 0.5226(16),\, 0.5230(17), \,0.5227(18)$ for $A_0=0.1\,\si{\volt}$ and $A_0=1.5\si{\volt},\,2.0\si{\volt},\,2.5\si{\volt}$, respectively.
The nominal value is in the range of $\Delta_{\text{nom}} = L_2/L_1 = 0.513 \, ...\, 0.551$ resulting from the precharacterization of inductors with $L_1 = 334\, ...\, 343\,\si{\micro \henry}$ and $L_2 = 176\, ...\, 184\,\si{\micro \henry}$ measured at $f=430\,\si{\kilo\hertz}$. The experimentally obtained values for the attenuation factor agree with the theoretical expectation of the linear limit.
We hence confirm that the LCn state inherits the spatial behavior, and thus its topological character, from the boundary state in the linearized SSH limit.
In contrast to the peak-to-peak voltage, the spatial profile of the RMS value, and accordingly the reactive power at each node, remains invariant upon the introduction of non-linearity (\supp~B). Locally at each node, the RMS value determines the parameter $m$ of the cnoidal wave solution in Eq.~\eqref{eq:single_resonator_solution} and traces all parameters back to the input.
Fig.~\ref{fig:evaluation}~(d) shows the spatial decay of the eccentricity $m$ towards the bulk obtained by solving the peak-to-peak voltage of the signal $u_\text{pp} = u(0)- u (T/2)$ for $m$. With larger voltage amplitudes close to the boundary of the circuit the influence of the non-linearity is stronger, and we expect a greater deformation of the wave. In agreement with Fig.~\ref{fig:measurement}, as $m$ increases towards the boundary, the peaks of the wave become sharper while the valleys get wider.
This is reflected in the attenuation factor of the exponential decay of $m$, which is measured to be $\Delta_{m} = 0.5249(25),\, 0.5221(33),\, 0.5209(65)$ for excitation amplitudes of $A_0 = 1.5\si{\volt},\,2.0\si{\volt},\, 2.5\si{\volt}$, respectively, and matches with the decay factor of $u_{\text{RMS}}$.

%-------------------------------------------------------------------------------

% ------------------------------------------------------------------------------
%  CONCLUSION
% ------------------------------------------------------------------------------

\paragraph*{Conclusion.}
Our one-dimensional periodic circuit network with Su-Schrieffer-Heeger type dimerization and on-site non-linearity exhibits an unprecedented topological voltage configuration, which we denote as the localized cnoidal state. The interplay of topological edge modes and non-linearity is probed within an experimental framework of exceptional accessibility and tunability.
Using a single-resonator approximation, we find that the Korteweg-de Vries equation with cnoidal wave solutions determines the waveform in the time domain. Intriguingly, the eccentricity, \ie~the deformation of the waveform due to the non-linearity, also decays exponentially from edge to bulk.
The topological character of the boundary mode in the strongly non-linear SSH regime suggests itself for further analysis, and might allow us to acquire a deeper understanding of non-linear topological matter.

%-------------------------------------------------------------------------------

% ------------------------------------------------------------------------------
%  ACKNOWLEDGEMENT
% ------------------------------------------------------------------------------

\paragraph*{Acknowledgement.}
The work is funded by the Deutsche Forschungsgemeinschaft (DFG, German Research Foundation) through Project-ID 258499086 - SFB 1170 and through the W\"urzburg-Dresden Cluster of Excellence on Complexity and Topology in Quantum Matter -- \textit{ct.qmat} Project-ID 390858490 - EXC 2147. T.He. was supported by a~Ph.D. scholarship of the German Academic Scholarship Foundation.
%-------------------------------------------------------------------------------

% ------------------------------------------------------------------------------
%  BIBLIOGRAPHY
% ------------------------------------------------------------------------------

% ------------------------------------------------------------------------------
%

\clearpage

\def\thesection{SUPPLEMENTAL MATERIAL~\Alph{section}}
\def\thesubsection{\Alph{section}.\arabic{subsection}}

\onecolumngrid
\begin{center}
	\textbf{\large Supplemental Material: \\ Observation of cnoidal wave localization in non-linear topolectric circuits}\\[.4cm]
 Hendrik Hohmann,$^{1}$, Tobias Hofmann,$^{1}$ Tobias Helbig,$^{1}$ Stefan Imhof,$^{2}$ Hauke Brand,$^{2}$ Lavi K. \\Upreti,$^{1}$ Alexander Stegmaier,$^{1}$ Alexander Fritzsche,$^{1}$ Tobias M\"uller,$^{1}$ Udo Schwingenschl\"ogl,$^{3}$ Ching\\ Hua Lee,$^{4}$ Martin Greiter,$^{1}$ Laurens W. Molenkamp,$^{2}$ Tobias Kießling,$^{2}$ and Ronny Thomale$^{1}$ \\[.1cm]
	{\itshape ${}^1$Institute for Theoretical Physics and Astrophysics,\\ University of W\"urzburg, D-97074 W\"urzburg, Germany\\
	${}^2$Physikalisches Institut, Universität Würzburg, 97074 Würzburg, Germany\\
	${}^3$Physical Sciences and Engineering Division (PSE),\\ King Abdullah University of Science and Technology (KAUST), Thuwal 23955-6900, Saudi Arabia\\
	${}^4$Department of Physics, National University of Singapore, Singapore, 117542}\\
	(Dated: \today)\\[1cm]
\end{center}
\twocolumngrid

\renewcommand{\theequation}{S.\arabic{equation}}
\renewcommand{\thefigure}{S\arabic{figure}}
\setcounter{equation}{0}
\setcounter{figure}{0}
% ------------------------------------------------------------------------------
%  Appendix A
% ------------------------------------------------------------------------------

\section{\\The Korteweg-De Vries equation}

In the ``monoatomic'' limit $L_1 = L_2 = L$ the nSSH model resembles the discrete version of a non-linear transmission line. If we consider a smooth excitation pattern, which extends over multiple unit cells, the continuum approximation can be used to recast the equations of motion (EOM) into the form of the Korteweg-de Vries (KdV) equation. Given a bell-shaped excitation inserted into a system governed by the KdV equation, a pulse soliton forms~\cite{FirstKdVCircuit1}. The wave peak travels through the circuit with constant shape and velocity, since the defocusing effect of the dispersion relation $\omega(k)$ is counteracted by the non-linearity.
In the case of a periodic input signal, e.g.~a sinusoidal waveform with a specific frequency, instead of a pulse, the non-linear transmission line sustains a cnoidal wave~\cite{doi:10.1080/147864495086207391, articleee1}.

\subsection{Derivation of the KdV equation from Kirchhoff's rules}

We label the nodes of the circuit by the index $n$ and denote the voltage measured at the $n$-th node with respect to ground by $V_n$. The non-linear capacitor (varicap diode) to ground is modeled as a voltage-dependent capacitance
\begin{align}
	C(V_n) = \dfrac{\text{d} Q_{C,n}}{\text{d} V_n} ,
\end{align}
where $Q_{C,n}$ is the charge of the varicap. Let $i_n$ be the current running through the inductors from node $n$ to $n+1$. Then fundamental Kirchhoff's circuit laws for the nSSH circuit read
\begin{subequations}
\begin{align}
	\label{eq:eom_1}
	L\, \dfrac{\text{d}\, i_{n}}{\text{d}t} &= V_{n-1}- V_{n}, \\
	\label{eq:eom_2}
	i_n &= \dfrac{\text{d} \, Q_{C,n}}{\text{d} t} + i_{n+1},
\end{align}
\end{subequations}
When we combine \eqref{eq:eom_1} and \eqref{eq:eom_2}, the EOM take the form
\begin{align}
\dfrac{\text{d}^2 \, Q_{C,n}}{\text{d} t^2} \,=\,\dfrac{1}{L}\,\biggl(V_{n-1} -2\,V_{n} +V_{n+1}\biggr).
\label{la1}
\end{align}
We expand $C(V)$ to linear order in the voltage around its operating point $V_{\text{offset}}$, \ie~$V_n = V_{\text{offset}} + v_n$,
\begin{align}
C(V_n) = C_0\,(1 - a_1\,v_n)
\end{align}
with $C_0 = C(V_{\text{offset}})$. The variable $v_n(t)$ denotes a small voltage signal around $V_{\text{offset}}$.
By substituting in the integral form of the charge $Q_{C_n} = \int_0^V C(V') dV'$, equation \eqref{la1} becomes
\begin{align}
\dfrac{\text{d}^2}{\text{d} t^2} \biggl(v_n - a v_n^2  \biggr)  = \dfrac{1}{L\,C_0}\,\biggl(v_{n-1} -2\,v_{n} +v_{n+1}\biggr),
\label{la2}
\end{align}
where $a = a_1/2$.\\
In order to find analytic solutions, several approximations are necessary. First, we apply the continuum approximation in the spatial dimension. This means, replacing the voltage at node $n$, $v_n(t)$, by a ``voltage field'' $U(x, t)$. This is possible when the voltage difference between subsequent nodes is small, i.e. the signal is considered as smooth. Next, one can Taylor expand the voltage field up to fourth order in $x$, which leads to the modified  Boussinesq equation~\cite{Doix1},
\begin{align}
	\dfrac{\partial^2 U}{\partial t^2} - c_0^2 \, \dfrac{\partial^2 U}{\partial x^2} - \dfrac{c_0^2}{12} \dfrac{\partial^4 U}{\partial x^4} - \epsilon a \dfrac{\partial ^2 U^2}{\partial t^2} = 0 .
\end{align}
Finally, we transform the equation into a moving frame of reference by $x \rightarrow x - c_0 t$, where $c_0 = 1/\sqrt{L C_0}$ is the phase speed of the linear system. This yields the KdV equation for a non-linear transmission line
\begin{align}
\dfrac{\partial U}{\partial t} + a\,c_0\,U\, \dfrac{\partial U}{\partial x} + \dfrac{c_0}{24} \, \dfrac{\partial^3 U}{\partial x^3} = 0.
\label{circuitKDV}
\end{align}
This equation for the ``voltage field'' dependent on space and time coordinates can be transformed to a dimensionless form by introducing the re-normalized variables
\begin{subequations}
\begin{align}
\tilde{t} &= t \, \dfrac{c_0}{3^{2/5}} , \\
\tilde{\Phi} &= U\, \dfrac{a}{3^{2/5}} , \\
\tilde{x} &= x\,\dfrac{6}{3^{4/5}} .
\end{align}
\end{subequations}
We recover the dimensionless standard KdV equation
\begin{align}
\dfrac{\partial \tilde{\Phi}}{\partial t} + 6 \tilde{\Phi} \, \dfrac{\partial \tilde{\Phi}}{\partial \tilde{x}} + \dfrac{\partial ^3 \tilde{\Phi}}{\partial \tilde{x}^3} = 0.
\label{std}
\end{align}

%-------------------------------------------------------------------------------
% Figure 1
%-------------------------------------------------------------------------------
\begin{figure}[]
	\centering
  \begin{tikzpicture}
  	\node[anchor=north west,inner sep=0pt] at (-0.07, 0)
		{
			\includegraphics[width=0.82\linewidth]{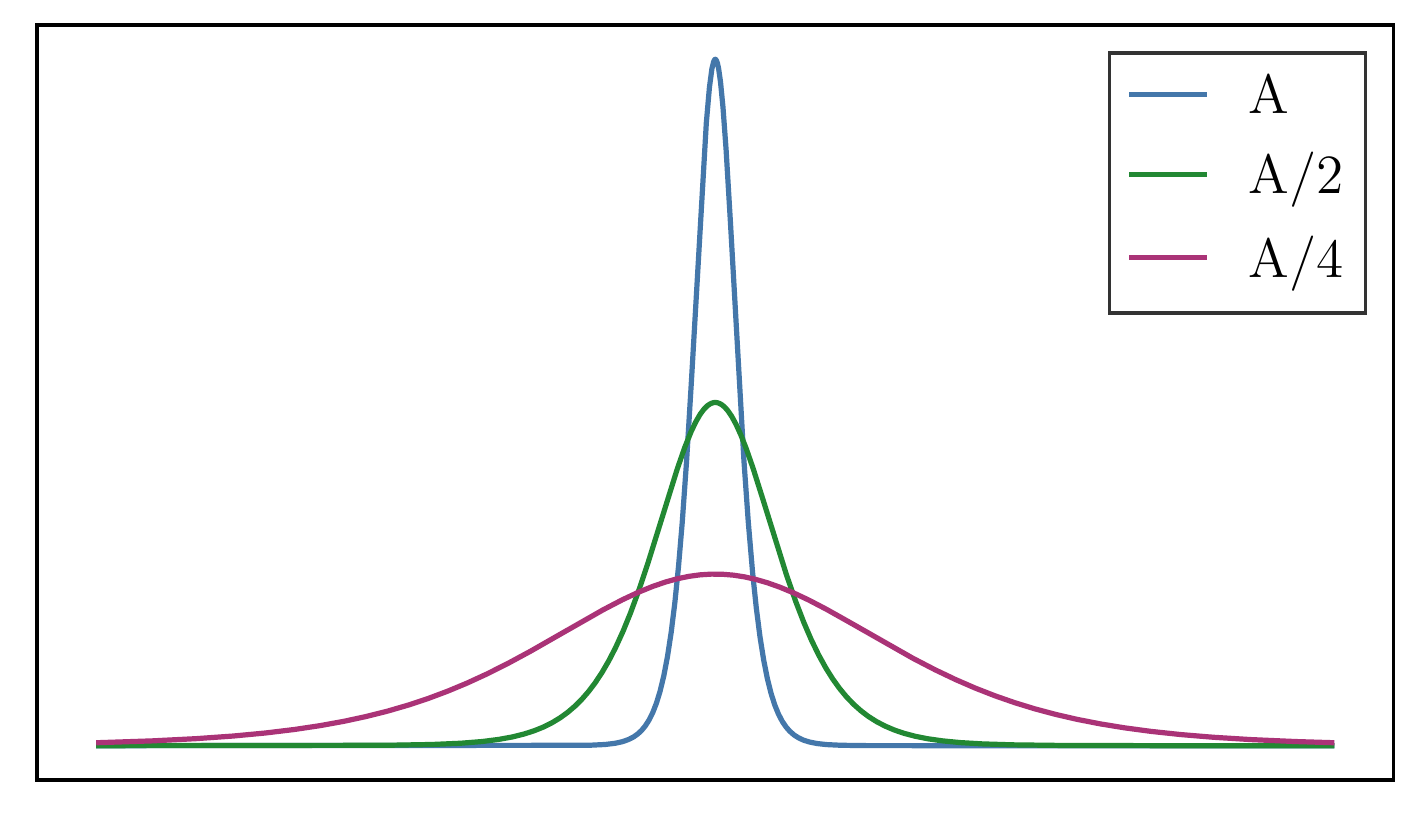}
		};
  	\node[anchor=north west,inner sep=0pt] at (0, -4.19)
		{
			\includegraphics[width=0.815\linewidth]{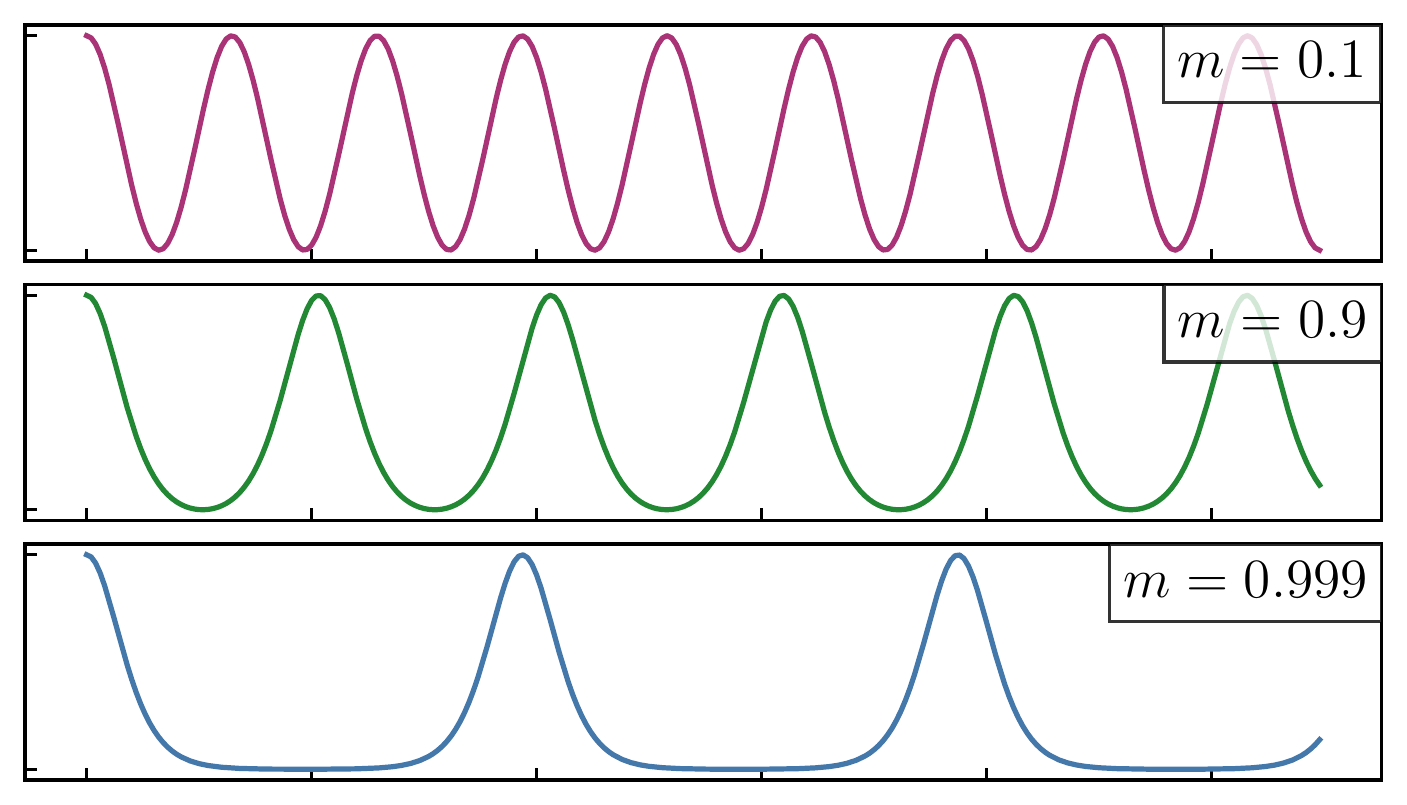}
		};
		\node[anchor=north east] at (0, 0) {(a)};
		\node[anchor=north east] at (0, -4.19) {(b)};
	\end{tikzpicture}
	\caption{(a)~Soliton solutions \eqref{solitons} for different amplitudes visualizing the correlation between width and amplitude.
(b)~Cnoidal wave solution \eqref{cnoidalw} for different elliptic parameters $m$. In the linear limit $m \rightarrow 0$, the Jacobi elliptic function cn$^2$ tends towards a cosine while for $m \rightarrow 1$ the periodic signal splits into separate solitons of sech$^2$ form \eqref{solitons}. 		\label{solandcn}
	}
\end{figure}
%-------------------------------------------------------------------------------

\subsection{Solutions of the KdV equation}

The non-linear, dispersive partial differential KdV equation~\eqref{std} was initially derived to describe waves in shallow water canals.
Different analytic solutions, namely soliton and cnoidal waves, were found by \mbox{Korteweg} and de Vries \cite{doi:10.1080/147864495086207391}. These solutions have the form
\begin{subequations}
\begin{align}
\tilde{\Phi}_{\text{Soliton}}(\tilde{x},\tau)
=& \dfrac{A}{2}\,\text{sech}^2\biggl(\dfrac{\sqrt{A}}{2}(\tilde{x}-A \tau) \biggr),\label{solitons} \\
\tilde{\Phi}_{\text{cn-wave}}(\tilde{x},\tau)
=& \tilde{\Phi}_2 + H\, \text{cn}^2 \biggl(2\, \text{K}(m) \,\dfrac{\tilde{x}- c \tau}{\lambda} \, \biggr| \,m \biggr),
\label{cnoidalw}
\end{align}
\end{subequations}
where $A$ is the amplitude of a soliton. For the cnoidal wave, $H$ can be viewed as the actual amplitude while $\tilde{\Phi}_2$ is the minimum (wave trough) and $\tilde{\Phi}_2 + H$ the maximum (wave peak). The function $\text{cn}(z \,|\, m)$ denotes the Jacobi elliptic cosine function and K$(m)$ the complete elliptic integral of first kind. $m \in [0,1]$ is the elliptic parameter, which we will call ``eccentricity'', and relates to the $\textit{elliptic modulus}$ $k$ according as $m = \sqrt{k}$.

A visualization of solitons with different amplitudes can be seen in Fig.~\ref{solandcn}(a). The cnoidal wave solution is displayed in Fig.~\ref{solandcn}(b) for different eccentricities. For small $m$ (linear limit) the wave tends towards a cosine function while separated soliton pulses are recovered for $m \rightarrow 1$.

%-------------------------------------------------------------------------------
% Figure 2
%-------------------------------------------------------------------------------
\begin{figure*}[]
\centering
    \begin{tikzpicture}
        \node[inner sep=0pt, anchor=west] at (-5.6, 3)
				{
					\includegraphics[width=0.287\linewidth]{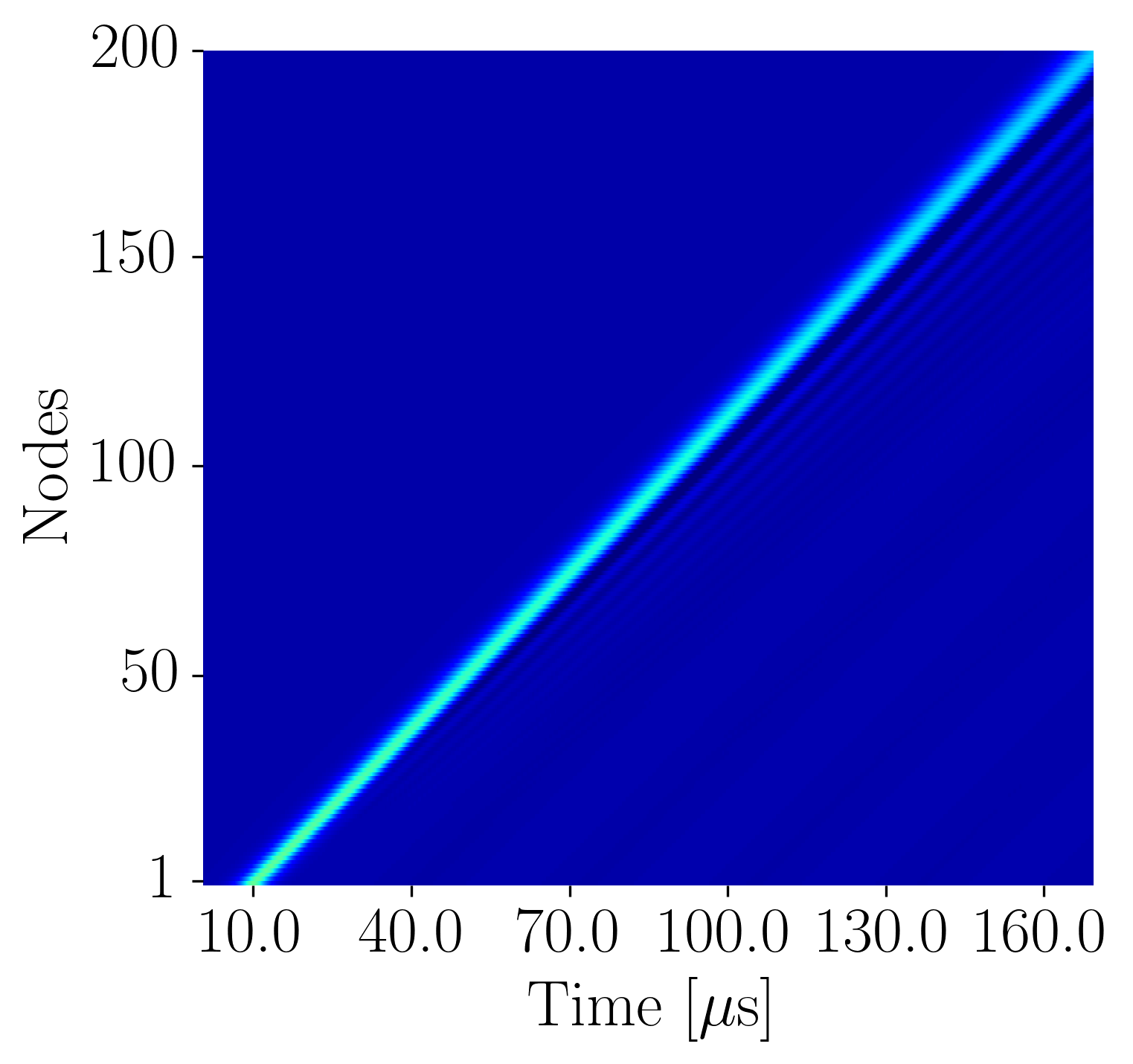}
				};
        \node[inner sep=0pt, anchor=west] at (-0.1, 3)
				{
					\includegraphics[width=0.325\linewidth]{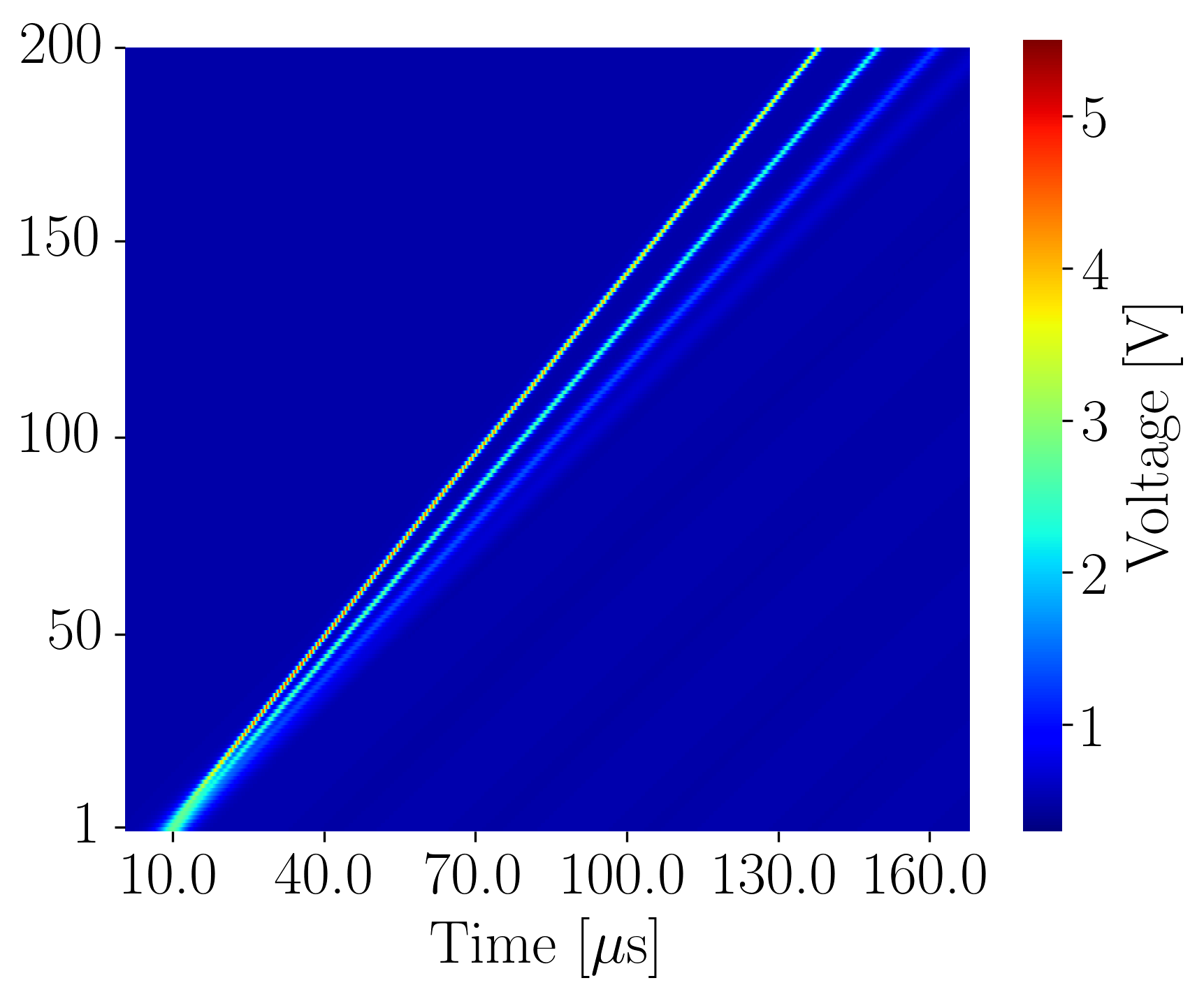}
				};
        \node[inner sep=0pt, anchor=west] at (6., 2.94)
				{
					\includegraphics[width=0.32\linewidth]{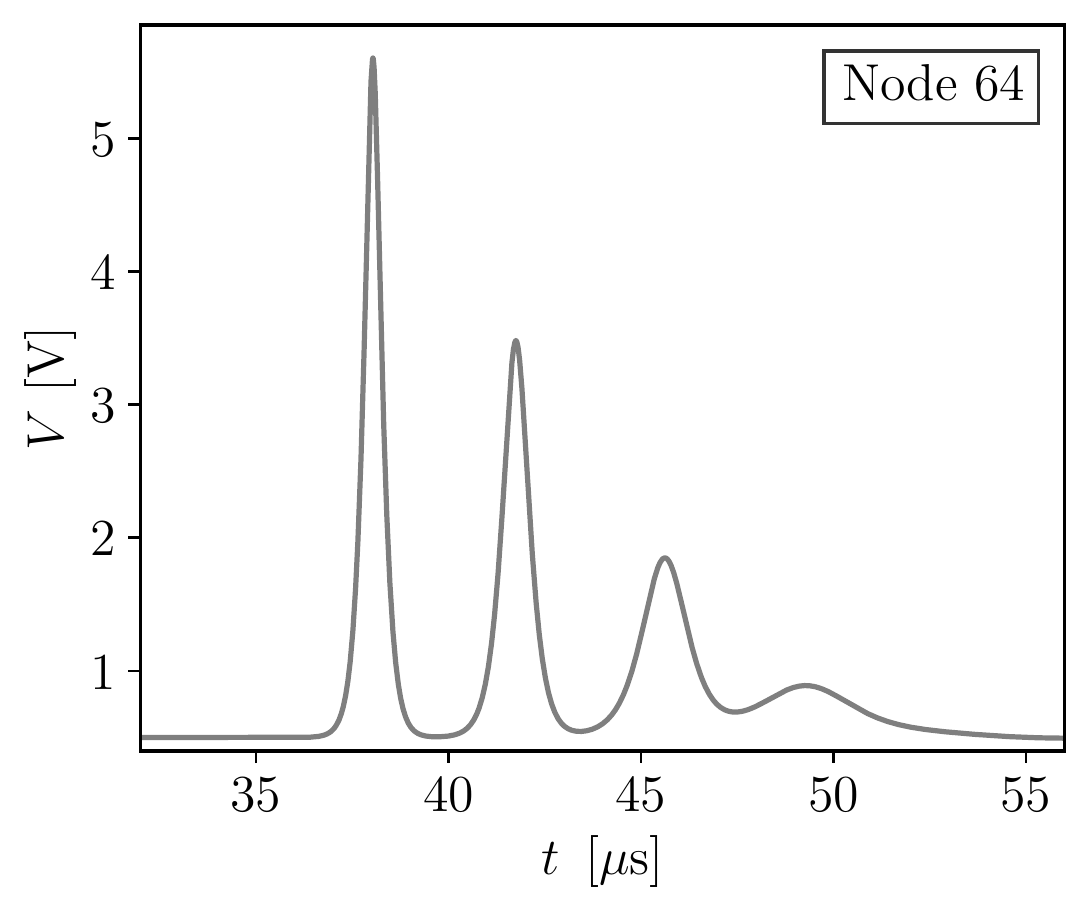}
				};
				\node[anchor=west] at (-5.6+0.7, 5.4)  { (a) Linear model};
        \node[anchor=west] at (0+0.4, 5.4)  { (b) Non-linear model};
        \node[anchor=west] at (6.5+0.05, 5.42)  { (c) };
			\end{tikzpicture}
\caption{Realistic LTSpice simulations of linear- (a) and non-linear transmission line (b,c) with $250$ nodes. A gaussian input is fed into node 1, the pulse propagates throughout the line, subjected to a small dispersion and dissipation. b) The same pulse as in (a) leads to a formation of solitons in the non-linear line. c) Cross section of the signal in (b) at node $64$, the three highest pulses possess a soliton like shape.}
\label{circuitpulse}
\end{figure*}
%-------------------------------------------------------------------------------

As it was shown in previous works, the non-linear electric circuit arrays can be described by the KdV equation with solitons and cnoidal wave solutions \cite{article1, FirstKdVCircuit1, Kofan1988,PhysRevE.49.8281}.
As a primarily test in the design stage of our circuit experiment, we performed realistic LTSpice simulations of the non-linear transmission line and compared them to the linear reference model as well as the theoretical expectation. Figure~\ref{circuitpulse} shows the simulation results if the (a) linear and (b) non-linear transmission line is excited with a Gaussian pulse. While the pulse disperses in case (a)  it splits into supersonic (\ie$c_\text{sol}>c_0$) solitons in case (b). Figure~\ref{circuitpulse}~(c) displays the temporal waveform of the splited solitons for one voltage node.

%-------------------------------------------------------------------------------
% Figure 2b
%-------------------------------------------------------------------------------
\begin{figure}
	\centering
	\includegraphics[width=0.9\linewidth]{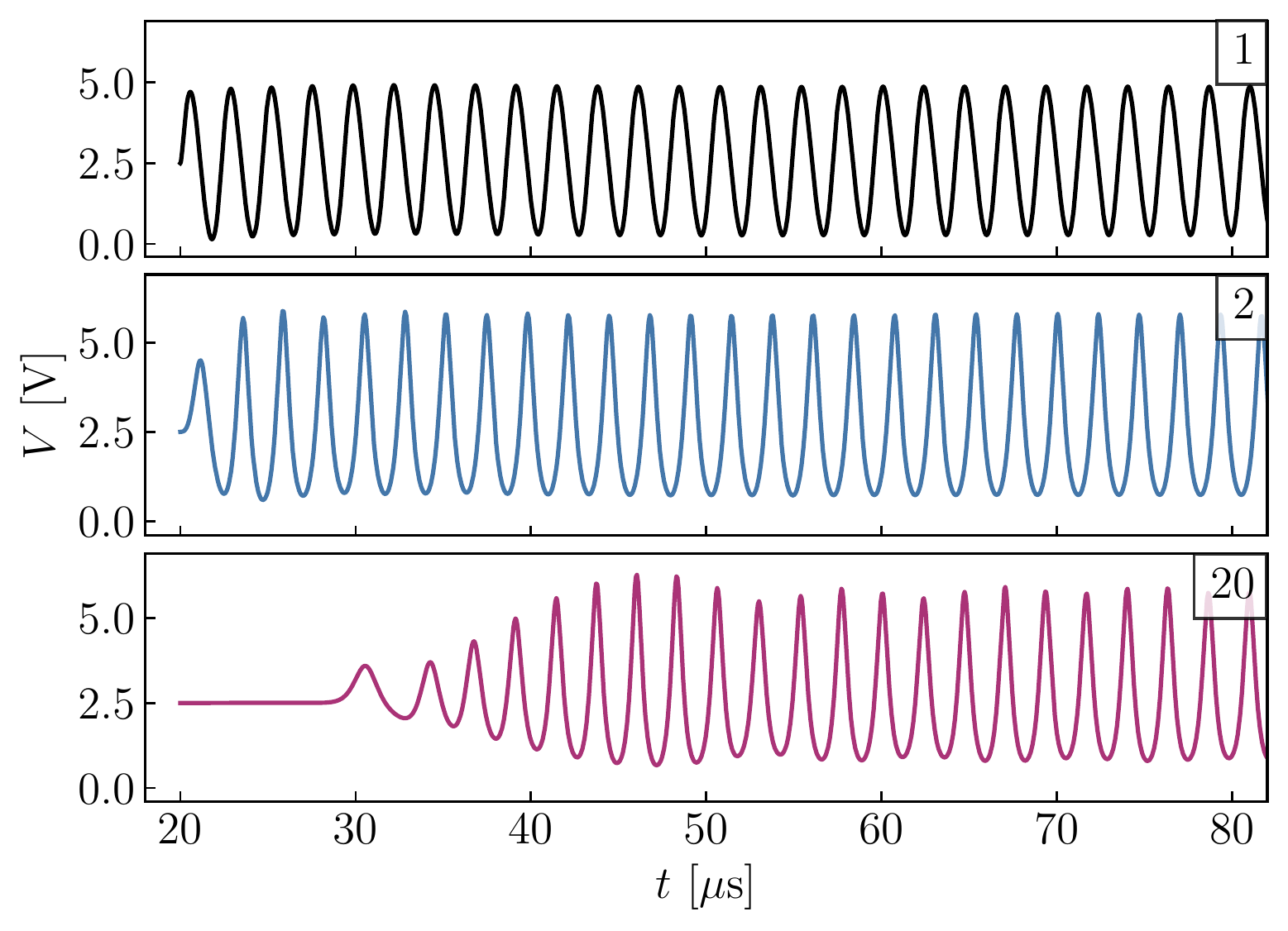}
	\caption{Voltage response measured in a realistic LTSpice simulations of a non-linear transmission line with $250$ nodes.  A constant sinusoidal input is fed into node $1$ of the non-linear transmission line, the input is deformed into a cnoidal wave. The excitation at node 1 suppresses the deformation.
		\label{fig:ltspice_sinusoidal}
		}
\end{figure}
%-------------------------------------------------------------------------------

When inserting a constant sinusoidal input at node 1, the formation of a cnoidal wave can be observed (see Fig.~\ref{fig:ltspice_sinusoidal}). For voltage nodes 2 and 20, we clearly recognize the deformation of the sine-shaped input signal into a cnoidal wave. For voltage node 1, however, this behavior is obstructed, since the voltage source is directly connected to this node. We keep this in mind for the analysis of the experimental results.

% ------------------------------------------------------------------------------
%  Appendix B
% ------------------------------------------------------------------------------

\section{\\Spatial character of the LC\lowercase{n} state}

%-------------------------------------------------------------------------------
% Figure 3
%-------------------------------------------------------------------------------
\begin{figure*}[t!]
    \centering
    \begin{tikzpicture}
        \node[inner sep=0pt] at (-5.6, 0)
				{
					\includegraphics[width=0.309\linewidth]{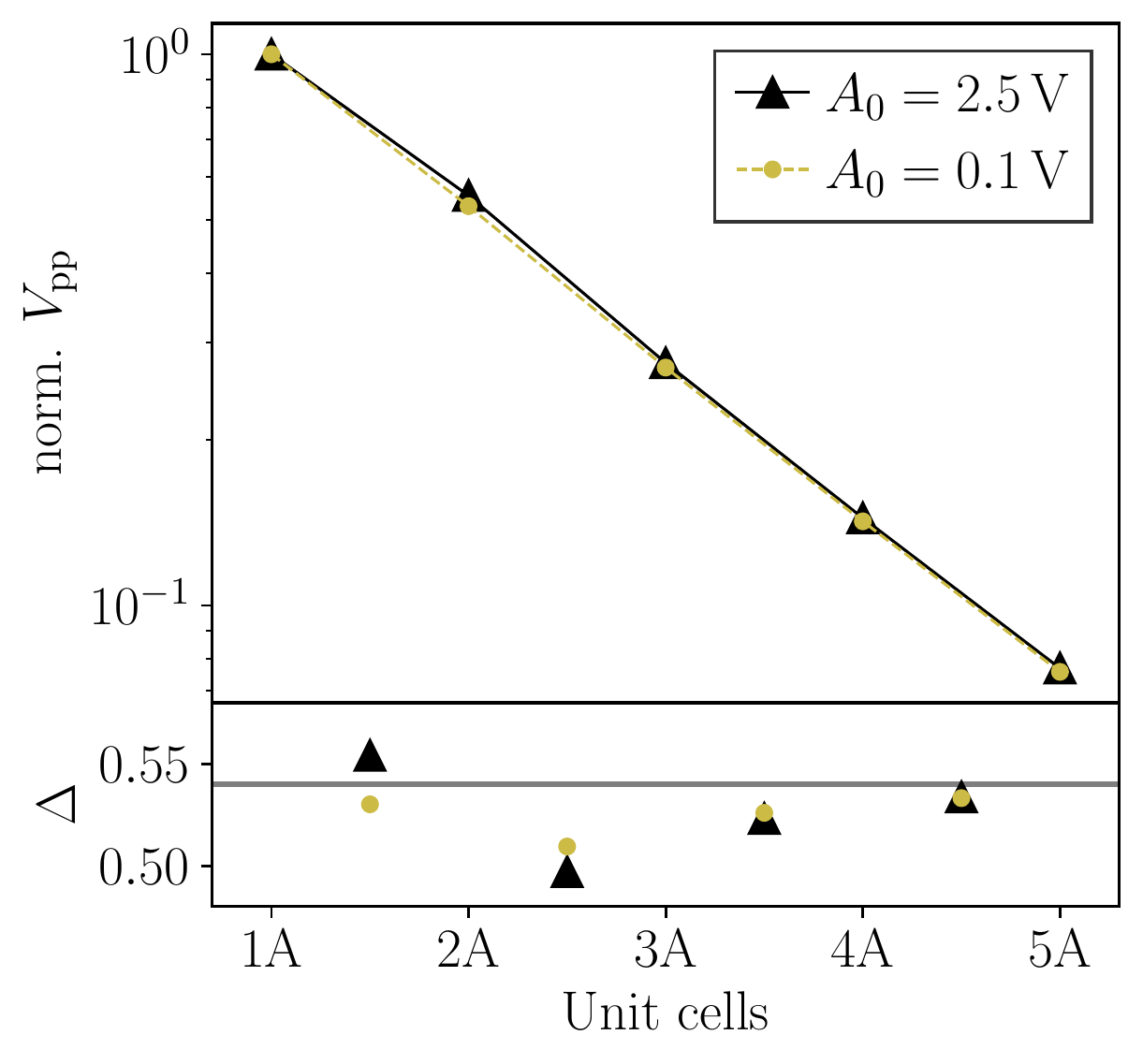}
				};
        \node[inner sep=0pt] at (0, 0)
				{
					\includegraphics[width=0.3\linewidth]{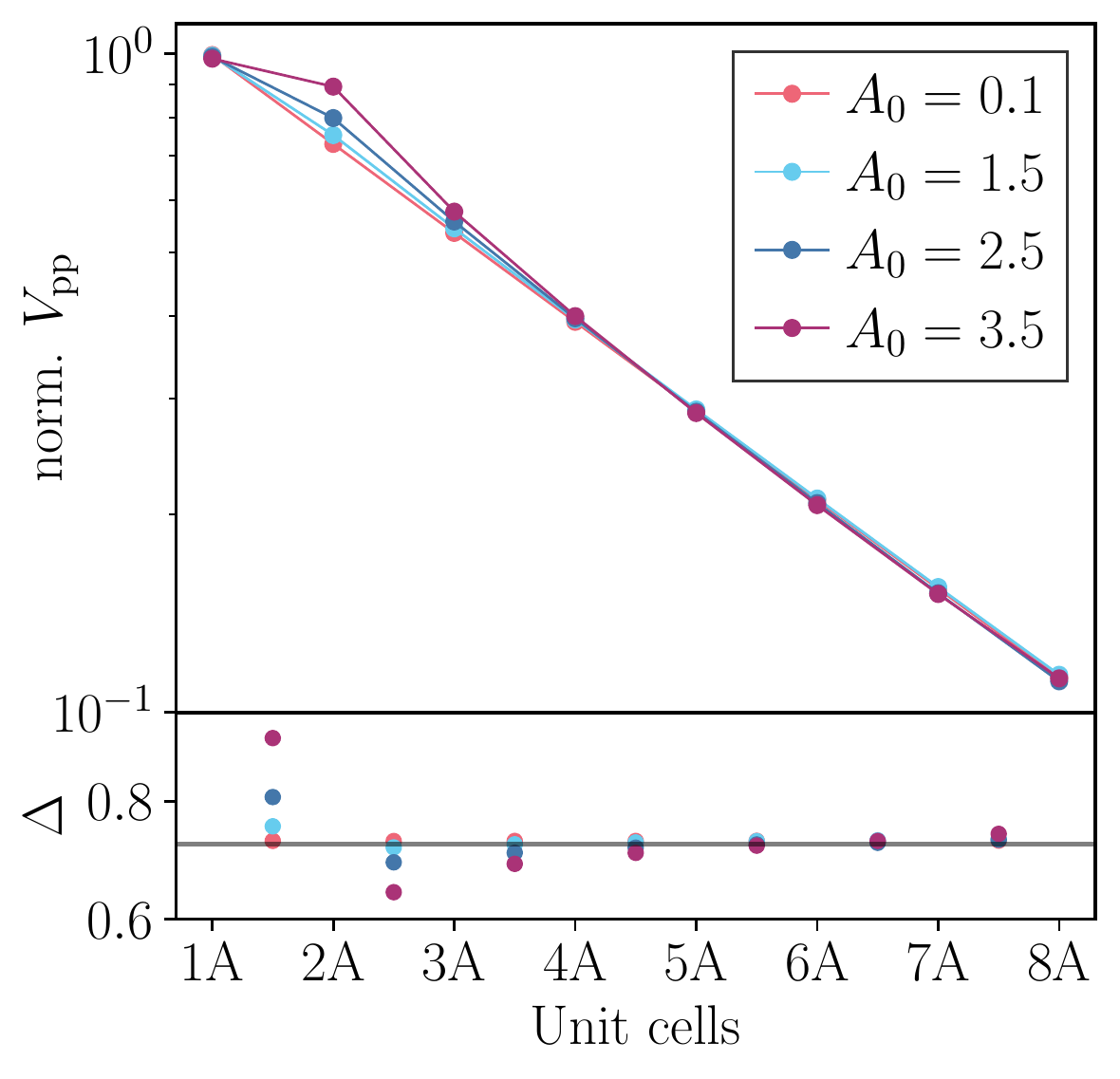}
				};
        \node[inner sep=0pt] at (5.6, 0)
				{
					\includegraphics[width=0.3\linewidth]{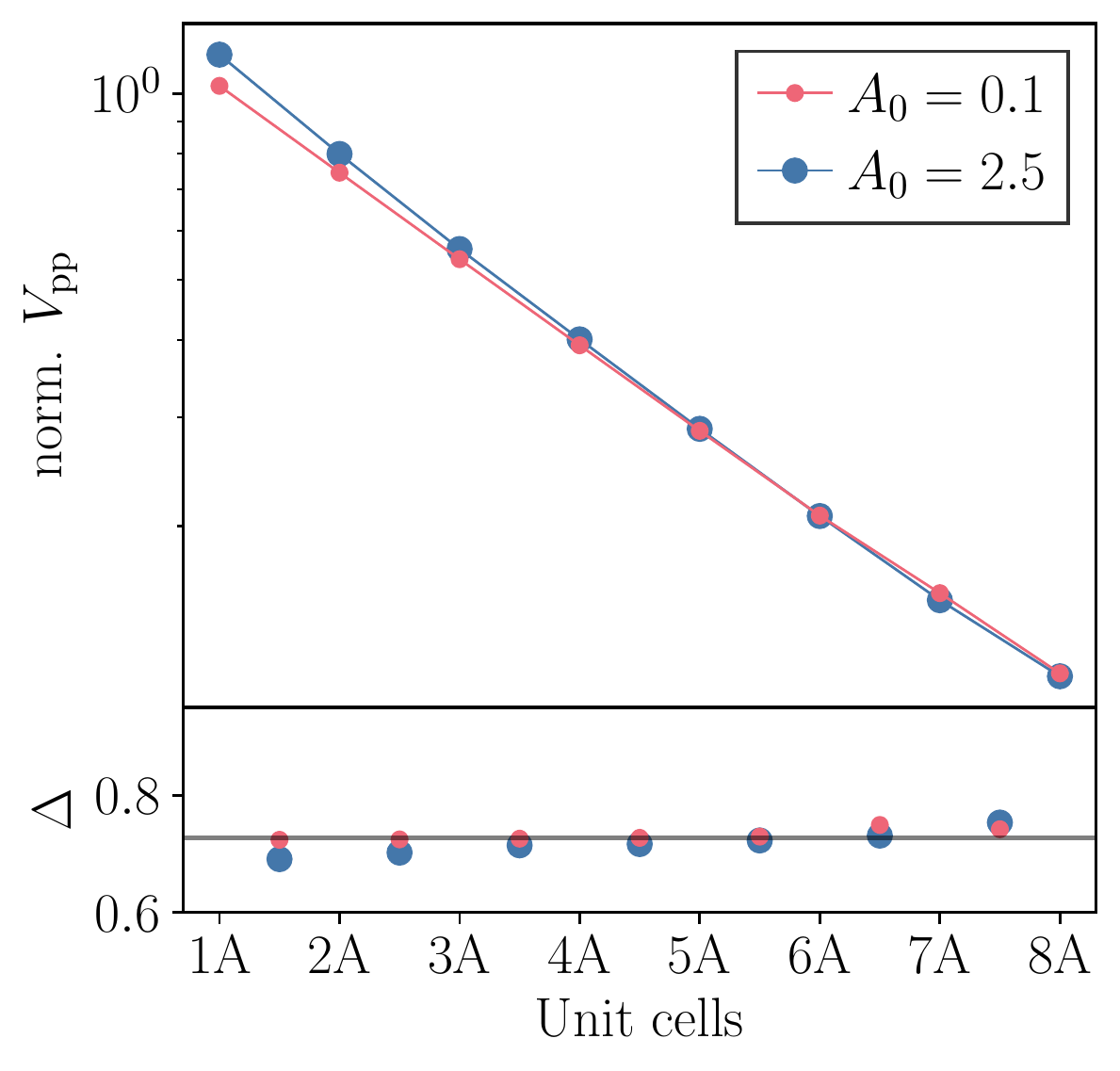}
				};
        \node[inner sep=0pt] at (-5.6, -5.35)
				{
					\includegraphics[width=0.309\linewidth]{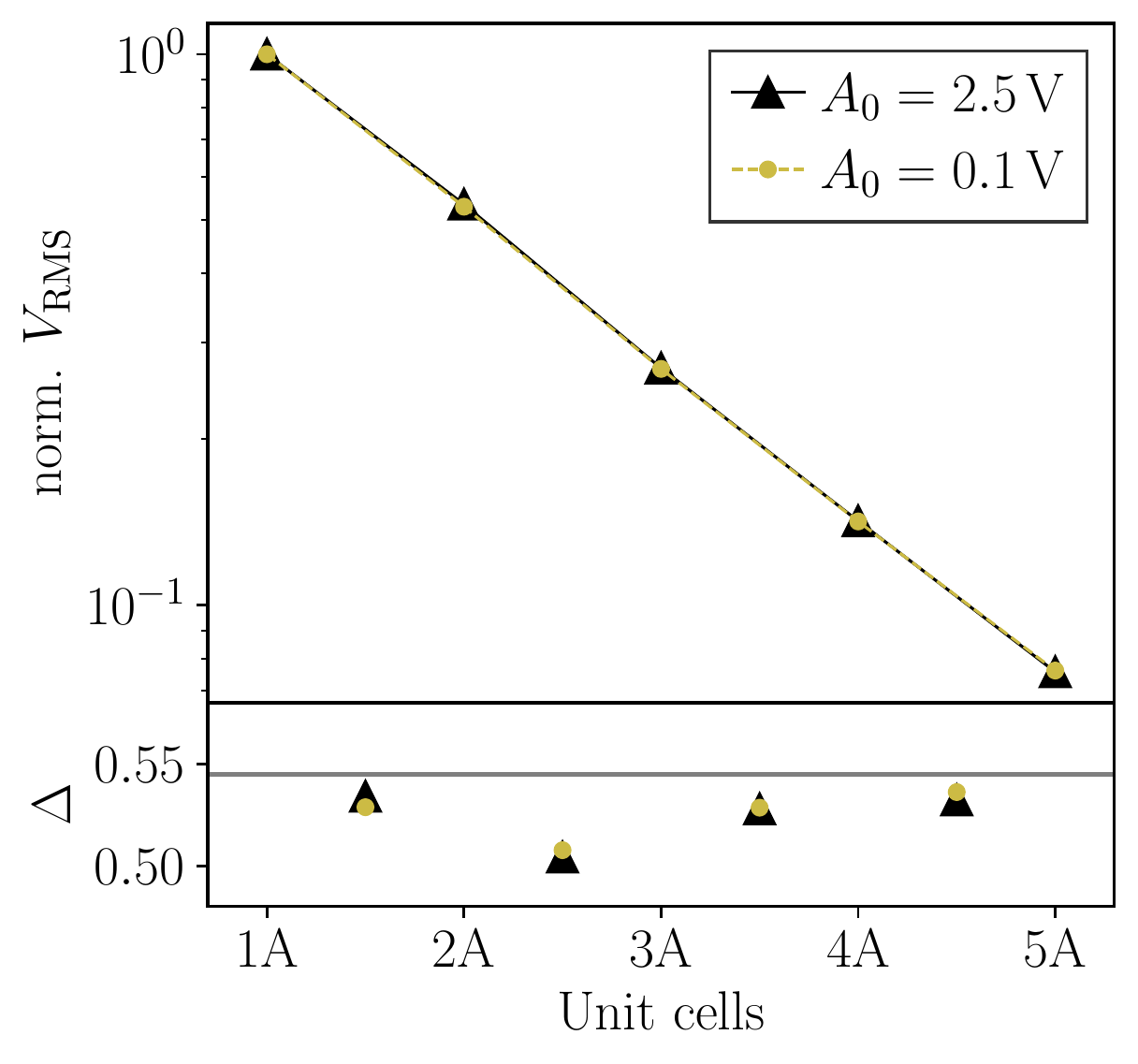}
				};
        \node[inner sep=0pt] at (0, -5.35)
				{
					\includegraphics[width=0.3\linewidth]{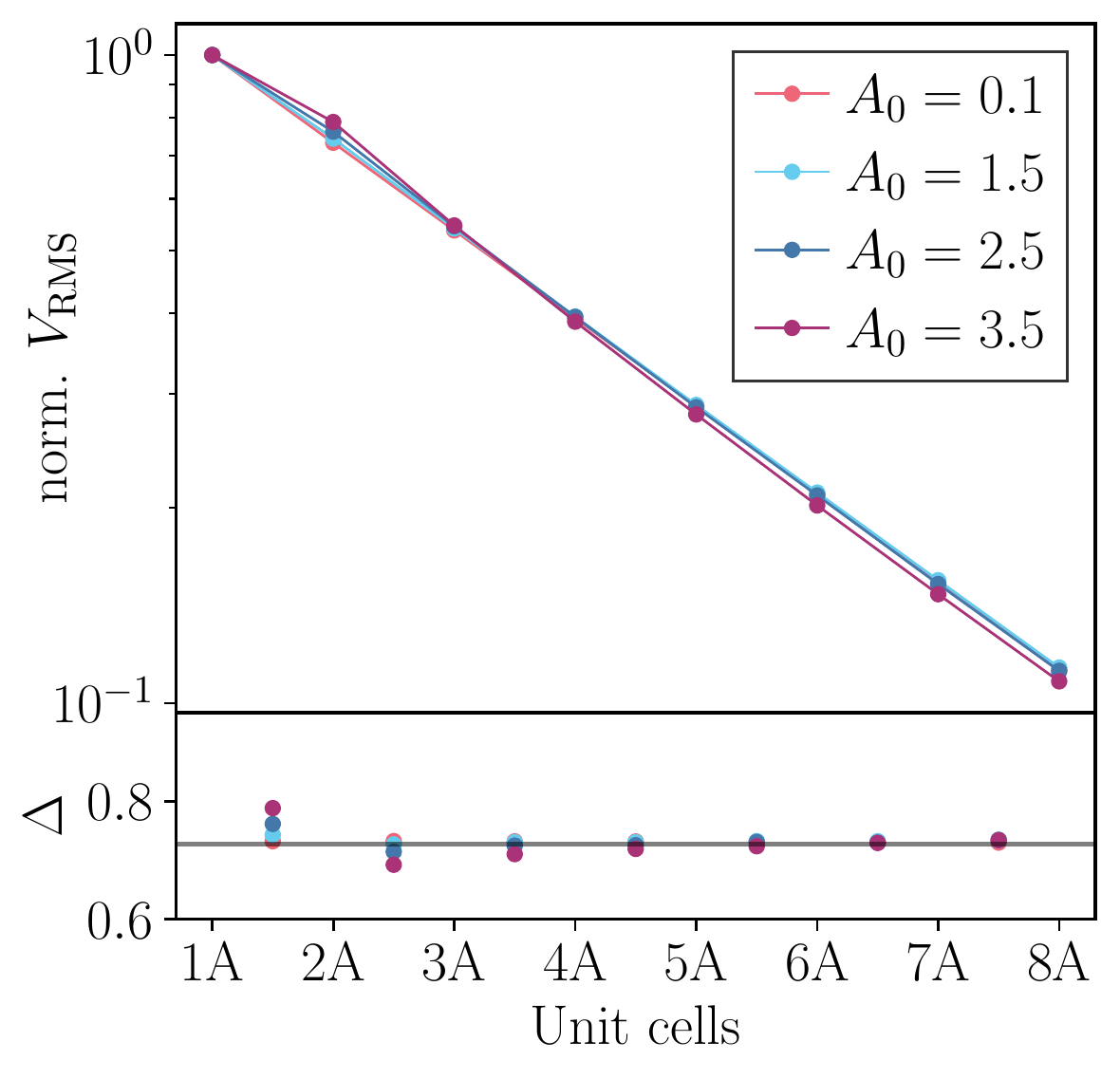}
				};
        \node[inner sep=0pt] at (5.6, -5.35)
				{
					\includegraphics[width=0.3\linewidth]{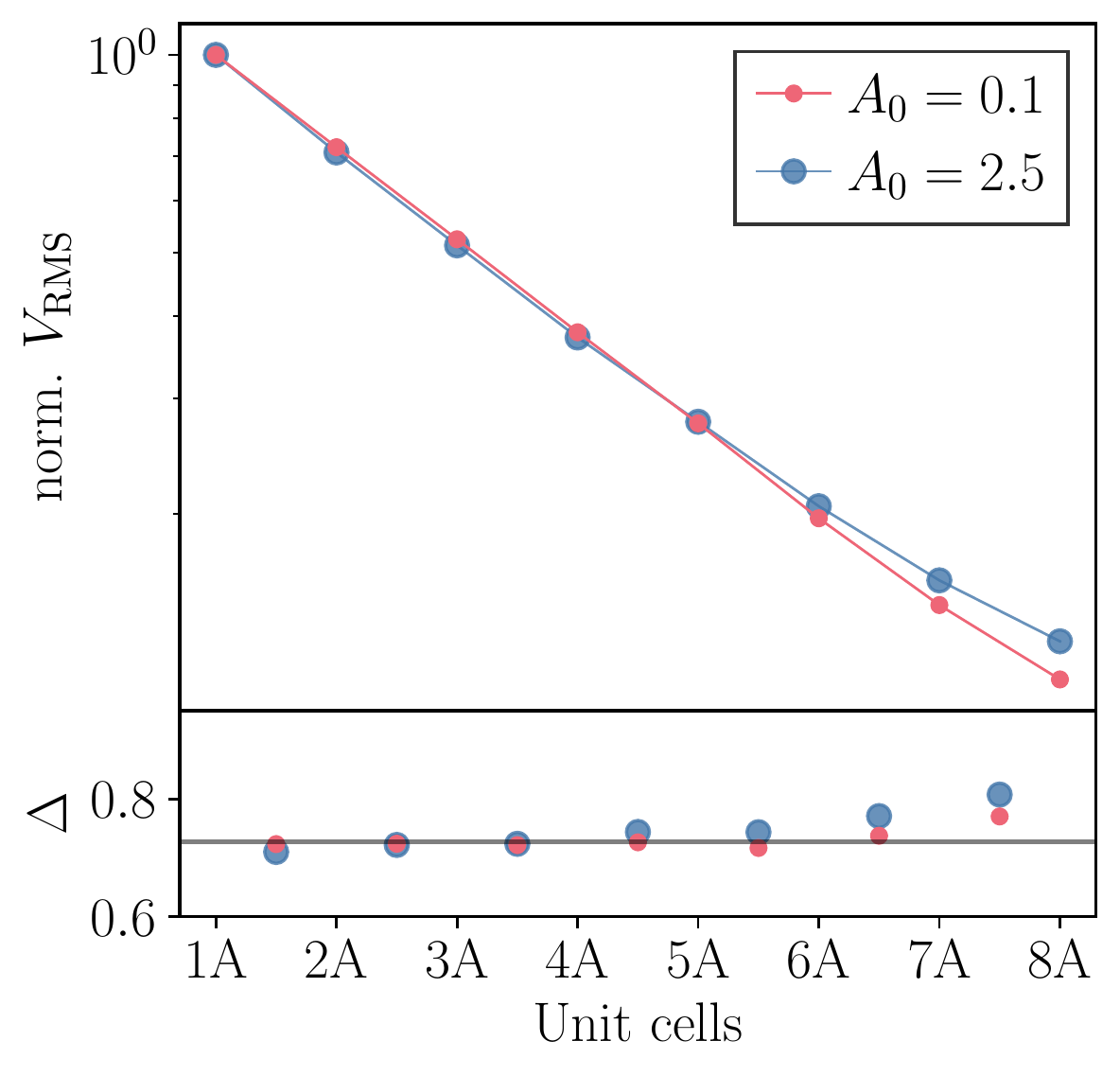}
				};

        \node at (-7.2-0.3, 2.7)  { (a) };
        \node at (-1.7-0.3, 2.7)  { (c) };
        \node at (3.9-0.3, 2.7)  { (e) };
        \node at (-7.2-0.3, -2.65)  { (b) };
        \node at (-1.7-0.3, -2.65)  { (d) };
        \node at (3.9-0.3, -2.65)  { (f) };
    \end{tikzpicture}
    \caption{
			The influence of on-site non-linearity on the spatial behavior of the SSH boundary state is illustrated, as a guidance to the eye the data points are connected by lines. In (a) and (b) the experimental measured data is shown ($L_{1,\text{nom}}=330$ and $L_{2,\text{nom}}=180$). While (a) shows the total amplitude of subsequent A nodes, (b) shows the quadratic mean square (RMS) amplitude of the signal. In contrast to the RMS the total amplitude is increased when non-linearities are excited. In realistic LTSpice simulation ((c), (d)) even higher non-linearities are obtained by choosing $L_{1,\text{nom}}=330$ and $L_{2,\text{nom}}=240$, as well as a higher $V_{\text{offset}}=3.5\,$V in a semi infinite line, this analysis confirms the trend in the experimental data. In (e) and (f) the same measurement was done including a perfect switch to decouple the line from the function generator.
    	\label{Fig:Localization}
		}
\end{figure*}
%-------------------------------------------------------------------------------

\subsection{Edge localization}

In this section we discuss some details on the LCn state regarding its spatial behavior, especially its localization properties (see suppl.~C for details on its temporal properties).

\paragraph*{Peak-to-peak vs. root-mean-square voltage.}

Figure~\ref{Fig:Localization} shows the exponential decay of the LCn state from the edge towards the bulk. Panels (a) and (b) display the measurement results, (c) to (f) depict simulation results for comparison.
In order to compare different input amplitudes, we normalize the voltages in all cases with respect to the input amplitude $A_0$.
In the lower part of the panels in Figure~\ref{Fig:Localization}, the ratio $\Delta = V_n / V_{n+1}$ of the amplitude on adjacent unit cells can be seen. The solid line represents the theoretically expected value as a reference.

In a non-linear circuit setup, the definition of the amplitude of a voltage signal is ambiguous: the peak-to-peak (P-P) voltage is defined as the difference between minimum and maximum in a period of the waveform (see Fig.~\ref{Fig:Localization} (a), (c), (e)), while the root-mean-square (RMS) value (see Fig.~\ref{Fig:Localization} (b), (d), (f)) averages the squared absolute voltage over one period and can be related to the power consumed by the circuit.

Consider the ratio plots (lower parts) of panels (a) and (b) or (c) and (d), respectively. One can clearly see that the spread between points belonging to different input amplitudes (and therefore different strengths of the non-linearity) is larger for the P-P as for the RMS values. The reason is that the P-P value depends strongly on the eccentricity of the waveform, while the RMS value as an average is less influenced by the deformation.
This means that for our purposes, the RMS is better suited as a measure for the excitation of a voltage node compared to the P-P value, in particular if the quantity is used to compare to the linear case.

Note, that the described discrepancy between P-P and RMS voltages is not a flaw in the experimental setup but a property induced by non-linearity, since the simulation results show equivalent behavior (see Fig.~\ref{Fig:Localization}~(c),~(d)). By choosing different values for the inductors $L_1$ and $L_2$ and a higher voltage offset, even a larger range of input amplitudes up to $A_0 = \SI{7,0}{\volt}$ could be investigated in the simulations. Regarding the differences between P-P and RMS voltage, the simulations agree with the experimental data.
\paragraph*{Free and driven LCn state.}

We like to investigate the influence of the voltage source, which is connected to the first node (1A) of the circuit on the localization properties of the state. In the experimental setup as well as in standard simulation the voltage source excites the first node permanently with a sinusoidal input current (applied voltage over shunt resistor). In order to check, whether or not the permanent driven influences the localization, we equipped our LTspice simulations with a perfect switch, which decouples the voltage source after some excitation time from the circuit. The results for the localization are depicted in Figure~\ref{Fig:Localization}~(e) and (f), respectively.\\
Since node 1A is directly connected to the voltage source, the cnoidal character of the voltage signal is obstructed at that node, as we already saw for the ``monoatomic'' chain and the KdV solution in Appendix~A. This effect slightly influences both P-P and RMS voltage values on the first node, and causes therefore also deviations in the first value of $\Delta$. As the simulation results in Figure~\ref{Fig:Localization}~(e) and (f) demonstrate, the deviation is absent if the voltage source is decoupled.
Moreover, the simulation shows that apart from node 1A the voltage values and ratios are barely influenced by the connected voltage source.

\subsection{Sublattice sturucture}

%-------------------------------------------------------------------------------
% Figure 4
%-------------------------------------------------------------------------------
\begin{figure*}[t]
	\centering
	\begin{tikzpicture}
  	\node[inner sep=0pt] at (-6.2, 0.06)
		{
			\includegraphics[width=0.371\linewidth]{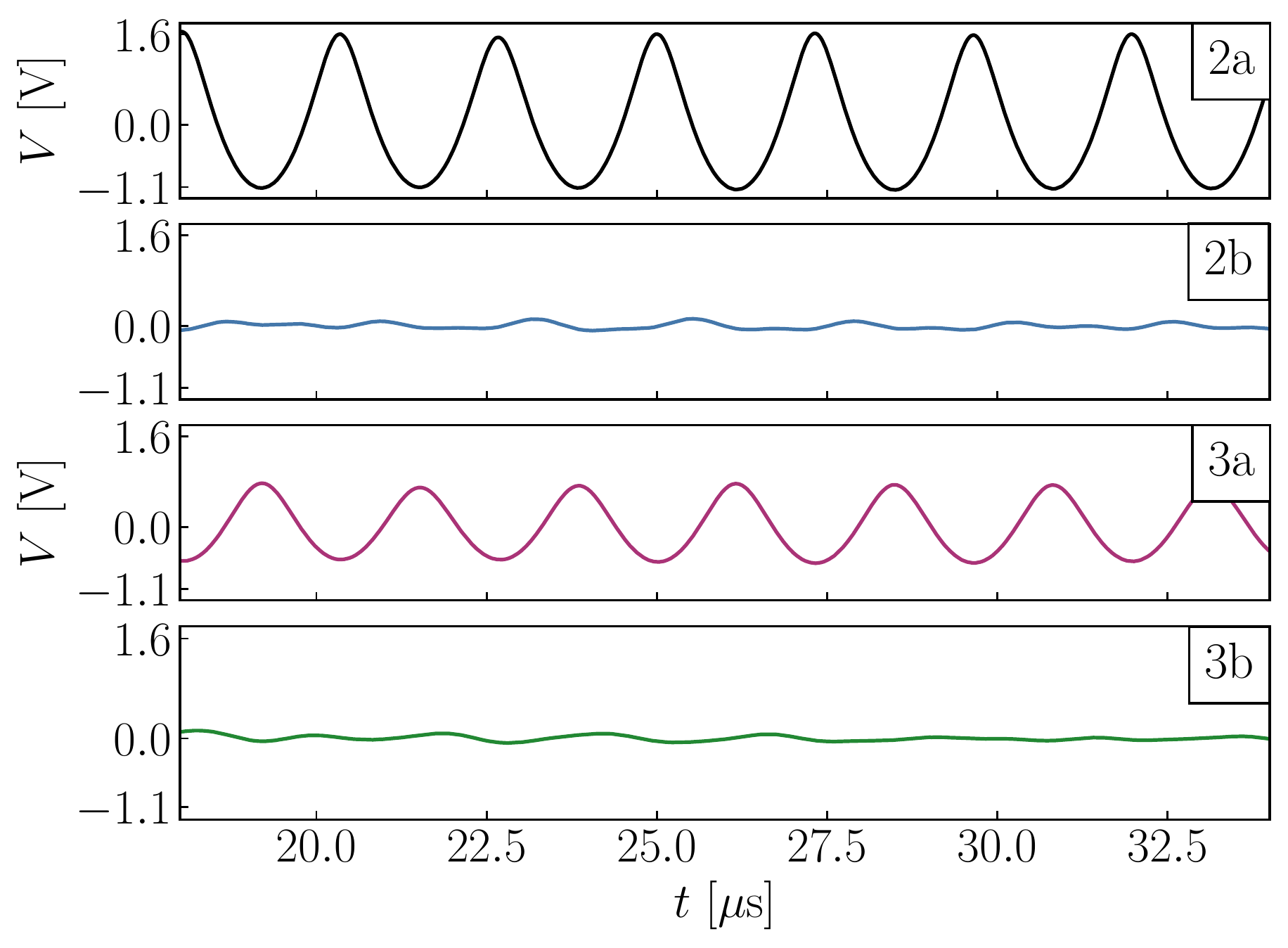}
		};
    \node[inner sep=0pt] at (0, 0)
		{
			\includegraphics[width=0.3\linewidth]{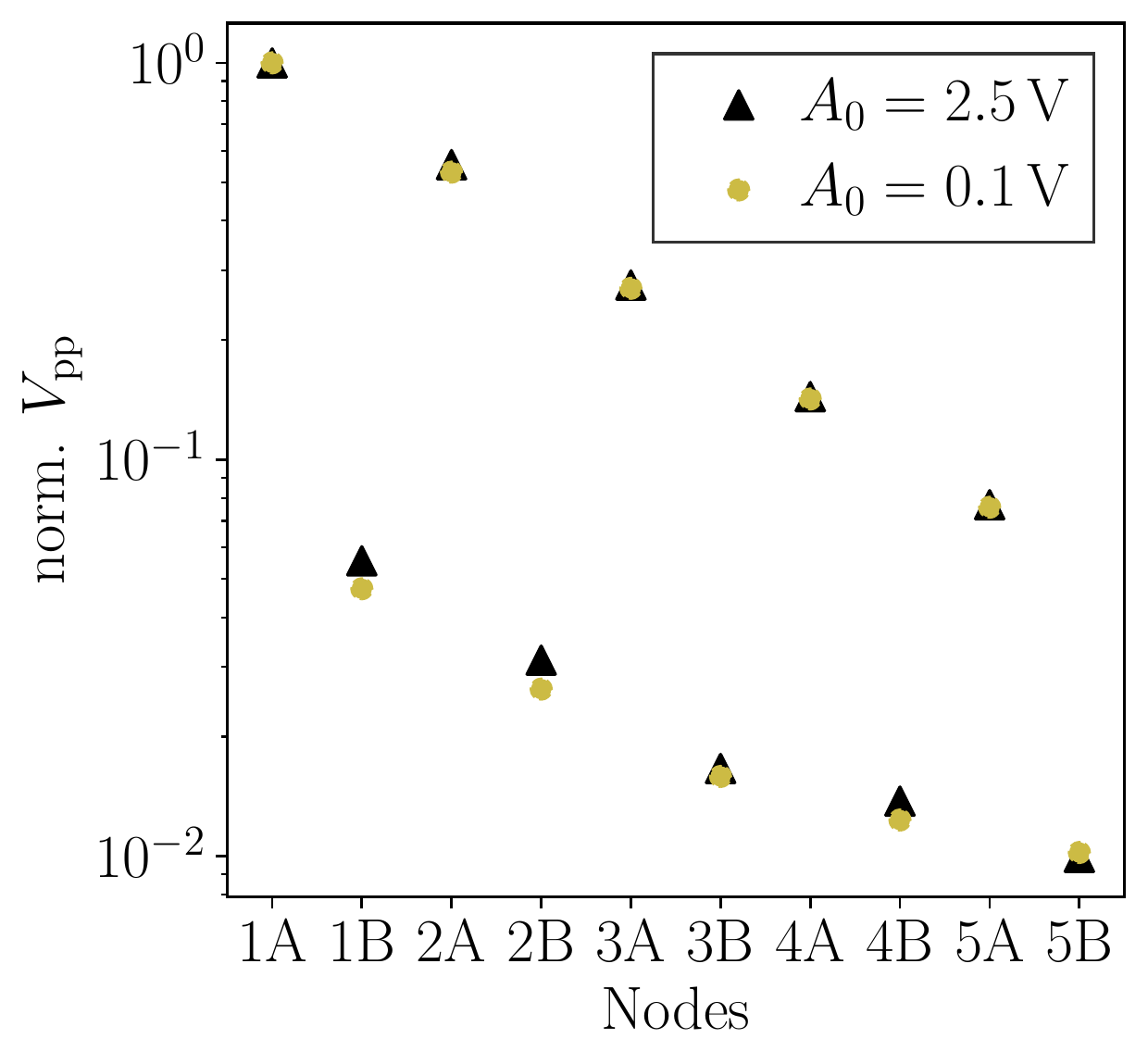}
		};
    \node[inner sep=0pt] at (5.6, 0)
		{
			\includegraphics[width=0.3\linewidth]{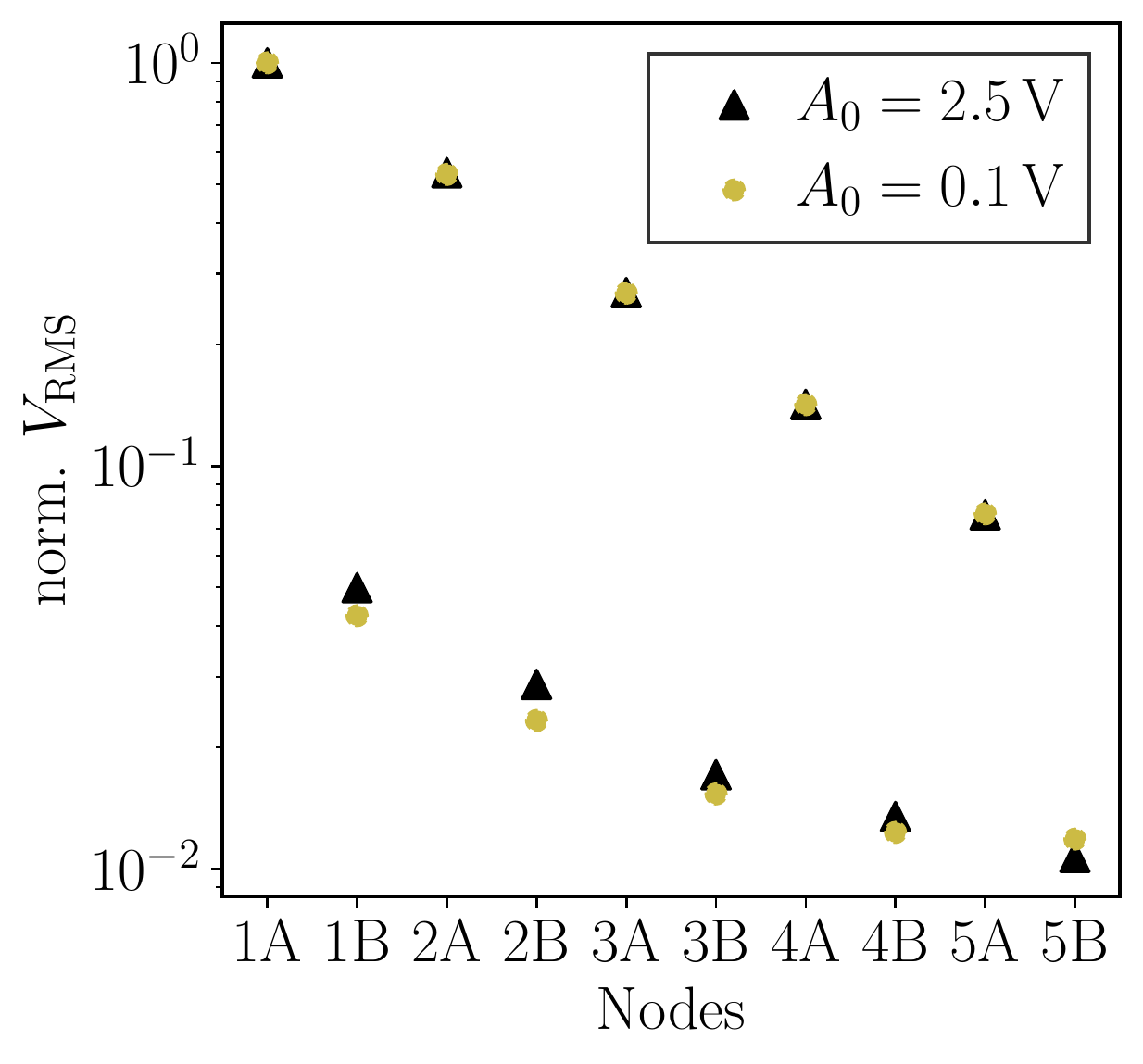}
		};
    \node at (-8.2-0.3, 2.65)  { (a) };
    \node at (-1.5, 2.65)  { (b) };
    \node at (4.1, 2.65)  { (c) };
	\end{tikzpicture}
	\caption{
		Depiction of measured sublattice character of the LCn state. (a) Nodes $2\text{A}- 3\text{B}$ of the measurement with $A_{0} = 2.5\,$V, due to the measurement procedure the signals are centered around $0\,$V even though $V_{\text{offset}}=2.5\,$V, see section D. The signal at sublattice B is small compared to A, indicating that the sublattice character of the linear SSH state remains intact in non-linear systems. In (b) and (c) the absolute amplitude and RMS normalized w.r.t. $1$A are shown for the linear limit $A_{0} = 0.2\,$V and non-linear system $A_{0} = 2.5\,$V. For both, total amplitude and RMS signals at B are about a magnitude smaller than the signal at subsequent A nodes. The effect of non-linearity is small compared to the absolute values within the observed eccentricity/amplitude range.
		\label{Fig:SubB}
	}
\end{figure*}

In order to verify that the LCn state inherits also the sublattice structure of the SSH edge state, we directly compare sublattice A and B of the first five unit cells.

\paragraph*{Validity of the single resonator approximation.}

The measured steady state LCn signal at nodes $2$A, $2$B, $3$A, and $3$B, for an input amplitude of $2.5\,$V (non-linear regime), can be seen in Figure~\ref{Fig:SubB}~(a).
As expected from the linear limit, the signal at sublattice B is small compared to neighboring A nodes. A quantitative analysis of amplitude and RMS values of the first 5 unit cells for $A_0 = \SI{2.5}{\volt}$ is displayed in Figure~\ref{Fig:SubB}~(b) and (c). All voltages are normalized with respect to node 1A. Additionally, we show the data in the linear limit ($A_0 = \SI{0.1}{\volt}$) as a reference.
We clearly observe that P-P and RMS voltages on sublattice B are approximately one magnitude smaller than the signals at the two neighboring A nodes. This validates the approximation of treating the voltage on sublattice B as negligibly small, and, thus, we can view nodes of sublattice B as virtual ground for the single resonator model. These findings are valid for both, the non-linear regime and the linear limit.

%-------------------------------------------------------------------------------
% Figure 5
%-------------------------------------------------------------------------------
\begin{figure}
	\centering
	\includegraphics[width=0.9\linewidth]{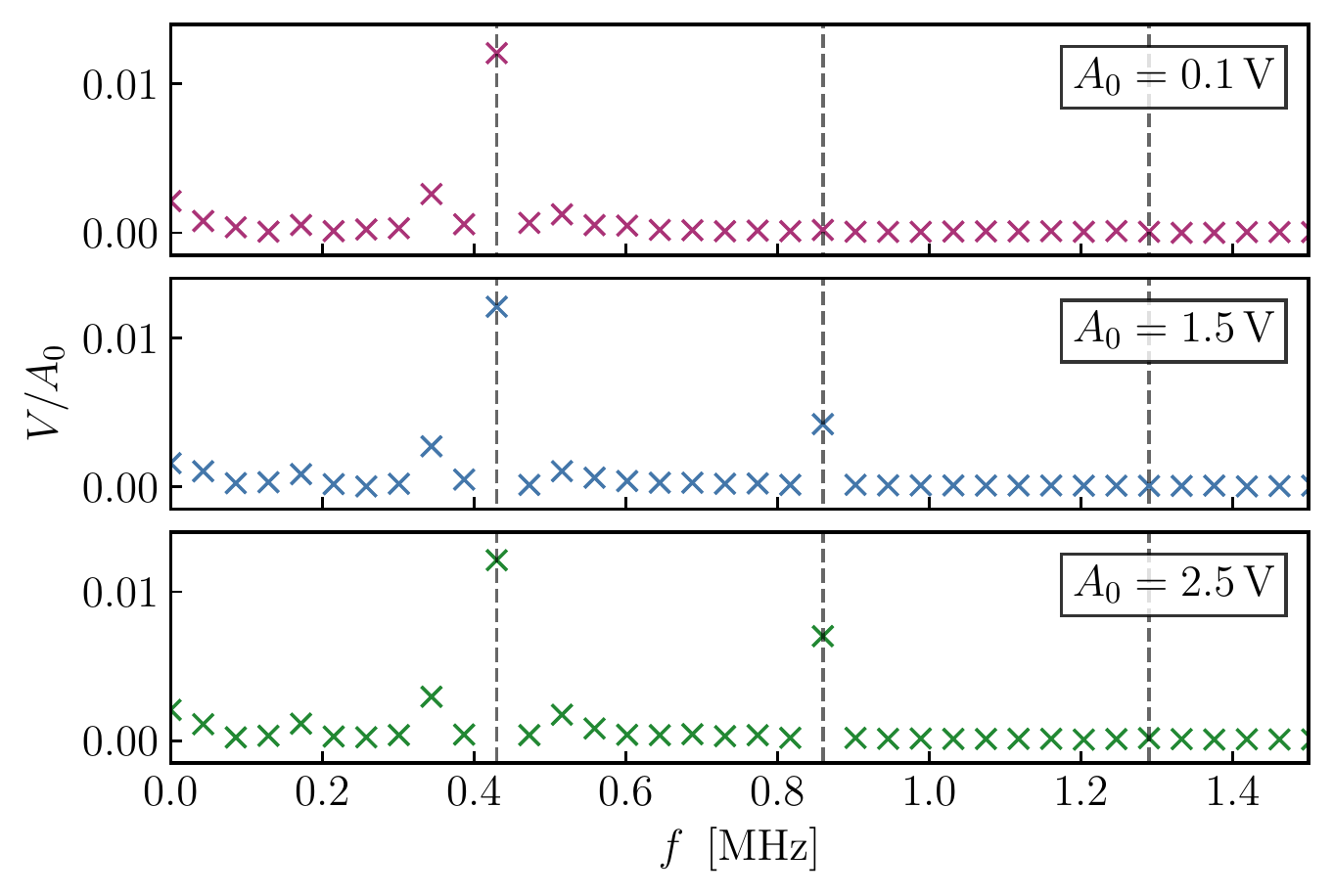}
	\caption{
		Fourier transformation of experimentally measured amplitude signal at node $2$B, normalized w.r.t. the input amplitude. The dashed lines indicate the resonance frequency $f_0$, the first, and second harmonic. For increasing amplitude the first harmonic is excited, indicating an additional coupling of the single resonators subjected to this frequency. The normalized magnitude of the peak at $f_0$ is observed to be constant.
		\label{FFTSubB}
	}
\end{figure}

\paragraph*{Frequency decomposition on sublattice B.}

In the main text we analyzed the frequency (Fourier) spectrum of the measured LCn state. It consists mainly of higher harmonics, \ie~integer multiples, of the basic excitation frequency $f_0$, and agrees with the theoretical predictions of the single resonator model.
The same analysis can be conducted on the small residual signal on sublattice B, which is shown in Figure~\ref{FFTSubB} at the example of node 2B. The figure depicts the Fourier spectrum for different input amplitudes, hence for different strength of the non-linearity.

First, we notice that the relative peak height at the base frequency $f_0$ is constant for different input amplitudes and thus independent of the eccentricity. This ensures that the overall signal on sublattice B does not become larger as the non-linearity increases, which is important for the single resonator approximation to be valid.

Moreover, the spectrum shows a peak at the first harmonic frequency $2\,f_0$ which grows with the non-linearity, but there is no excitation of the second harmonic $3\,f_0$. This means that in contrast to sublattice A, the waveform on sublattice B is not described by a $\text{cn}$-function and has not the cnoidal wave character.
Beyond that, the growing peak at $2\,f_0$ indicates an increased coupling between subsequent A nodes with increasing non-linearity.

% ------------------------------------------------------------------------------
%  Appendix C
% ------------------------------------------------------------------------------

\section{\\Single resonator model}

\begin{figure}[]
\centering
    \begin{tikzpicture}
        \node[inner sep=0pt] at (-2, 0.43)
{\includegraphics[width=0.375\linewidth]{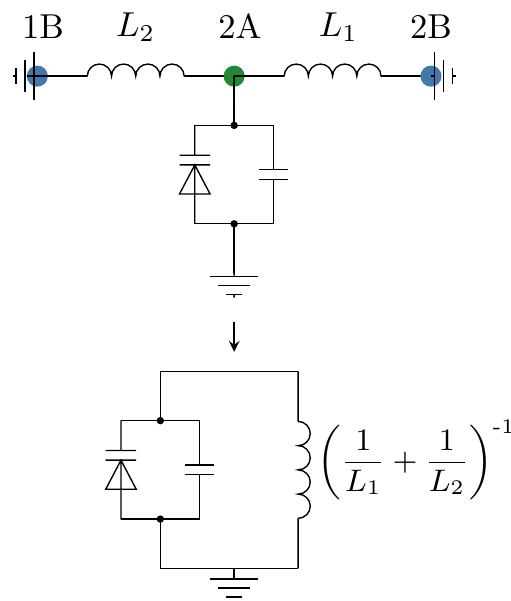}};
 \node[inner sep=0pt] at (2, 0.06)
{\includegraphics[width=0.25\textwidth]{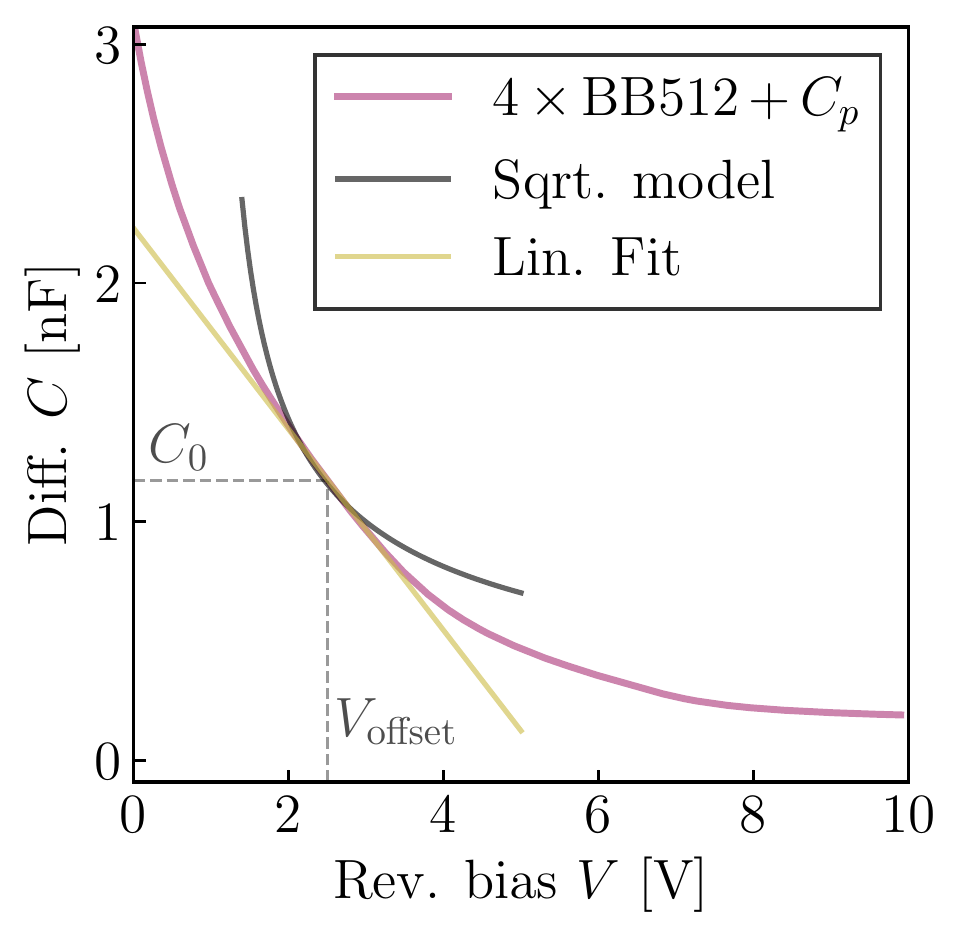}};
\end{tikzpicture}
\caption{(left) Transformation of a segment of the nSSH circuit to a non-linear LC resonator by treating the sublattice nodes B as virtual ground, due to their negligible amplitude in the LCn state. (right) Differential capacitance of 4 BB512 varicap diodes in a parallel circuit configuration complemented by an additional parasitic capacitance of the measurement setup. A fit of eq.~\eqref{c(v)} close to the bias voltage $V_{\text{offset}}$ is shown in black, along with a linear fit at $V_{\text{offset}}$ in yellow. }
\label{fig:singleresonator}
\end{figure}

In supplemental material B we confirmed that the signal at sublattice type B indeed vanishes for excitations around midgap frequencies. Therefore, we can treat the nodes of sublattice type B as AC-ground and decouple the circuit network into a chain of single resonators. In the AC analysis, we are only interested in the oscillatory (AC) part of the voltage signal and set the DC voltage offset $V_{\text{offset}}$ to the same potential as the AC-ground. Exemplified in Fig.~\ref{fig:singleresonator} for node $2$A, each single resonator can be recast as an LC resonator with inductance $L = (L_1^{-1} + L_2^{-1})^{-1}$ and non-linear capacitance $C(V)$. In the linear limit, its resonance frequency matches the resonance of the midgap SSH edge state, $f_0 = 1/(2 \pi \sqrt{L C_0} )$.
In summary, we approximate the nSSH circuit, excited at gap frequencies on the boundary and in the steady state, as a line of decoupled, unperturbed LC resonators with respective signal strengths according to their distance from the boundary, given by the SSH boundary state (cf. suppl. B). In this approach, the single resonators follow homogeneous differential equations, as they are decoupled from the external input at the boundary. Accordingly, any free parameter arising in the solution due to this decoupling is fixed by the voltage amplitude at the respective node. The voltage amplitude, in turn, is given by the real space voltage profile of the topological edge mode, which is scaled by the input amplitude. In the following, we show that the charge of the varicap diodes in the non-linear LC resonator follows the KdV equation~\cite{Blyuss2002ChaoticBO1, DeaneNonlinearDO1} with cnoidal waves as periodic solutions. Futhermore, we fix the eccentricity parameter $m$ of the cnoidal wave by matching its peak-to-peak amplitude with the experimental value at each node for a given input amplitude.

Applying Kirchhoff's second law to the closed circuit loop in Fig.~\ref{fig:singleresonator} we obtain
\begin{align}
V_C\,=\,-V_L \text{ with } V_L\,=\, L\, \frac{\text{d}^2Q}{\text{d}t^2},
\label{first}
\end{align}
where $V_C$ and $V_L$  are the voltage differences across the varicap-diode and the inductor respectively.
For voltages $V$ close to $V_{\text{offset}}$, we model the voltage dependent differential capacitance of the varicap by the equation for the junction diode, \ie~an inverse power law
\begin{align}
C(V)\,=\, \dfrac{\mathcal{C}}{\left(1+\frac{V-\nu}{\phi}\right)^{\gamma}}.
\label{c(v)}
\end{align}
Here, we use $V>0$ due to the reverse bias configuration and typical parameter ranges of $\gamma= 0.3 - 0.5$ and $\phi= 0.7 - 0.8$~\cite{PhysRevLett.47.13491,phdthesis1}. Deviating from the standard equation to model varicap diodes in electrical engineering, we introduce an additional degree of freedom ($\nu > 0$) in order to match the fit at voltages around $V_{\text{offset}}$.
A general derivation of the differential equation for arbitrary $\gamma$ can be found in~\cite{PhysRevLett.47.13491,phdthesis1}, whereas we employ $\gamma=0.5$ in the following, which results in the KdV equation for the charge on the capacitor.
Fig.~\ref{fig:singleresonator} shows the the differential capacitance from each node in the circuit to ground. Alongside, we plot a linear expansion around $V_{\text{offset}}=\SI{2.5}{\volt}$ and an inverse square root fit of equation~\eqref{c(v)} with $\gamma=0.5$, where the value of the capacitance and its slope are matched at $V=V_{\text{offset}}$. The linear expansion is used to derive the KdV equation for an electric transmission line in the continuum limit (cf. suppl. A).
In order to derive the differential equation of the single resonator, we use the inverse square root law model of equation~\eqref{c(v)} \ie~setting $\gamma=0.5$. With that, we obtain the fit parameters $\mathcal{C} \approx 2214.2\,$pF, $\phi \approx 0.4\,$V and $\nu \approx 1.44\,$V.

Varicap diodes with smaller capacitance values are in larger agreement with the model in equation \eqref{c(v)} over a broader voltage range. Nevertheless, for the experiment we choose the Siemens BB512 diode with its large capacitance, in order to obtain small eigen- and gap frequencies (cf. suppl. D). This ensures a high measurement precision in the experiment given the finite sampling rate of the oscilloscopes in the time domain. We conclude, that one of the major limitations in the theoretical treatment of the LCn state originates in deviations of the fitted square root model to the measured differential capacitance for voltages not close to $V_{\text{offset}}$.

%-------------------------------------------------------------------------------
% Figure 6
%-------------------------------------------------------------------------------
%-------------------------------------------------------------------------------

%-------------------------------------------------------------------------------
% Figure 7
%-------------------------------------------------------------------------------
\begin{figure*}[t!]
\centering
    \begin{tikzpicture}
            \node[inner sep=0pt] at (-2, 0.43)
{\includegraphics[width=0.9\linewidth]{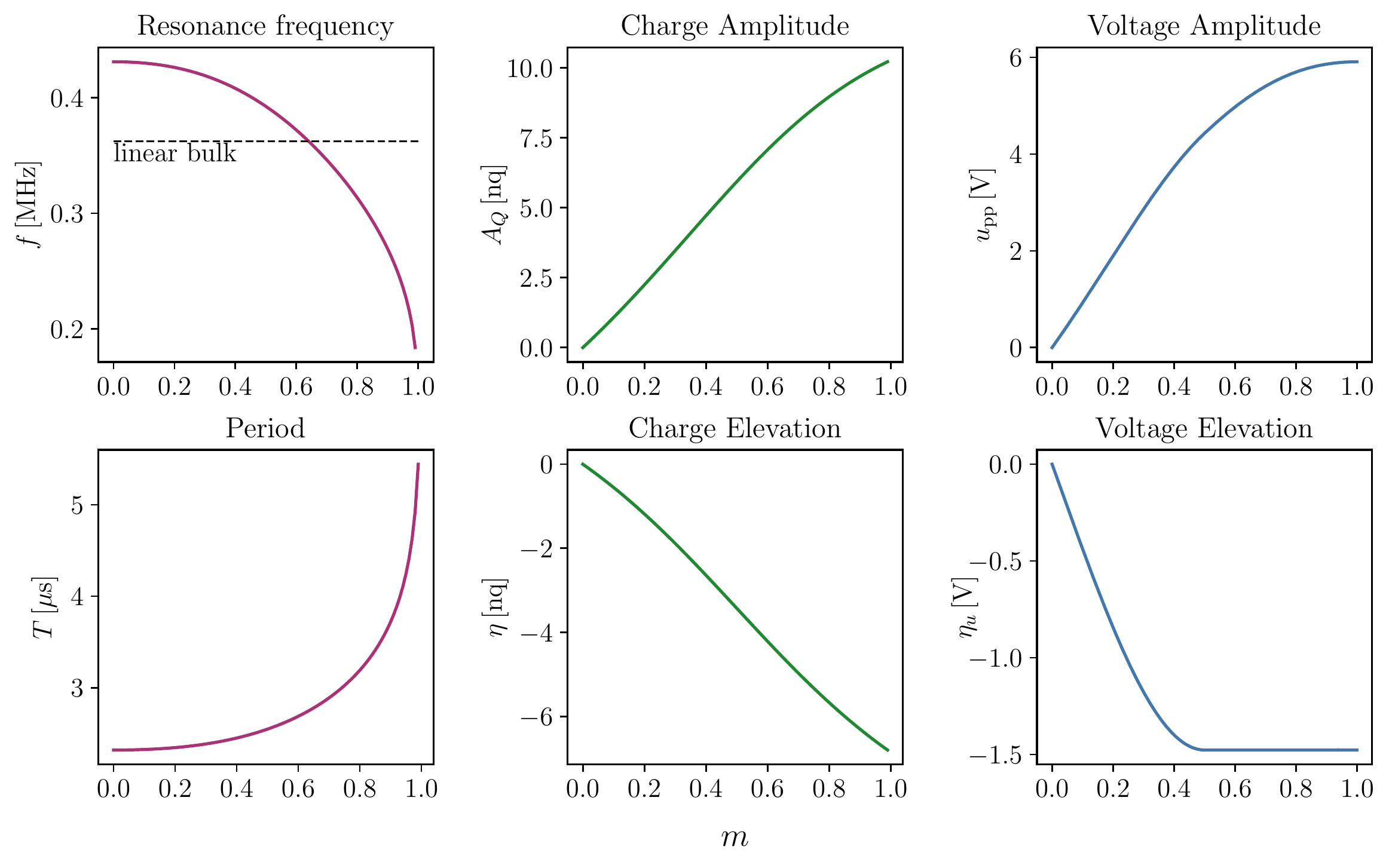}};
    \node at (-9.2-0.15, 5.2)  { (a) };
    \node at (-3.9, 5.2)  { (c) };
    \node at (1.3, 5.2)  { (e) };
        \node at (-9.2-0.15, 0.45)  { (b) };
    \node at (-3.9, 0.45)  { (d) };
    \node at (1.3, 0.45)  { (f) };
\end{tikzpicture}
\caption{Parameters of the single resonator approximation. Whereas the resonance frequency is the same for capacitor charge $Q$ and voltage signal, total amplitudes and elevations differ qualitatively. Voltage and charge elevation are relative to their offset values. The finite value of $\eta$ for $m \rightarrow 1$ indicates that the soliton limit is not recovered for $u_C(t)$.
}
\label{Parameters}
\end{figure*}

The charge of the capacitance is given by an integration over the differential capacitance
\begin{align}
Q(V_C(t))\,=\,\int_{0}^{V_C(t)} C(V)\, \text{d}V \text{ with } V_C(t)=V_{\text{offset}}+u_C(t),
\end{align}
where $u_C(t)$ is a small AC voltage around the DC operation point $V_{\text{offset}}$. The integral can be split into two contributions,
\begin{align}
Q(V_C(t))\,=\, \underbrace{\int_{0}^{V_{\text{offset}}} C(V)\, \text{d}V}_{=Q_{\text{offset}} =\text{const.}} + \int_{V_{\text{offset}}}^{V_{\text{offset}}+u_C(t)} C(V)\, \text{d}V,
\end{align}
the first corresponds to the charge due to the DC offset, whereas the second contains the AC contribution. We define the time-dependent part of the charge $\delta Q$ corresponding to the AC voltage contribution as
\begin{align}
\delta Q(u_C(t))\,&\equiv\, Q(V_C(t))-Q_{\text{offset}}\\
&=\,\int_{V_{\text{offset}}}^{V_{\text{offset}}+u_C(t)} C(V)\, \text{d}V.
\end{align}
where the offset charge is subtracted. A substitution $V \rightarrow V-V_{\text{offset}}$ leads to
\begin{align}
\delta Q(u_C(t))\,=&\,\int_{0}^{u_C(t)} C(V_{\text{offset}}+V)\, \text{d}V \nonumber\\\nonumber
 \,=&\,\mathcal{C} \,\int_{0}^{u(t)} \frac{1}{\sqrt{1+\frac{(V+V_{\text{offset}})-\nu}{\phi}}}\, \text{d}V
 \\ \,=&\, 2\, \mathcal{C} \,\phi \sqrt{1+\frac{(V+V_{\text{offset}})-\nu}{\phi}} \Biggr|^{u_C(t)}_0 \nonumber \\
  \,=&\,2\, \mathcal{C}\, \phi \biggr[ \sqrt{1+\frac{u_C(t)+V_{\text{offset}}-\nu}{\phi}} \nonumber\\
   &\,\,\,\,\,\,\,\,\,\,\,\,\,\,\,\,\,-\sqrt{1+\frac{V_{\text{offset}}-\nu}{\phi}}  \biggl].
\end{align}
By solving for $u_C(t)$ we obtain a non-linear relation between charge on and voltage across the non-linear capacitance,
\begin{align}
u_C(t)\,=\,\dfrac{\delta Q^2+ \delta Q \,4\, \mathcal{C} \, \phi \sqrt{1+\frac{V_{\text{offset}}-\nu}{\phi}}}{4 \,\mathcal{C}^2 \, \phi}.
\label{U(Q(T))}
\end{align}
Inserting $u_C(t)$ into equation \eqref{first} yields the non-linear differential equation for the charge on the varicap due to the AC signal
\begin{align}
L\, \dfrac{\text{d}^2\delta Q}{\text{d}t^2} \, =\,-u_C(t)\,=\,-\frac{\delta Q^2+ \delta Q\, 4 \,\mathcal{C} \, \phi \sqrt{1+\frac{V_{\text{offset}}-\nu}{\phi}}}{4\, \mathcal{C}^2 \, \phi}. \nonumber
\end{align}
We further define \mbox{$\Theta = 2\, \mathcal{C}\, \phi \sqrt{1+ \frac{V_{\text{offset}}-\nu}{\phi}} > 0$} and \mbox{$\beta= 2\, \mathcal{C}^2\, \phi\,L >0 $} as system dependent parameters of the differential equation, which results in the compact form
\begin{align}\label{eq:DEQcompact}
\ddot{\delta Q} + \frac{1}{2 \beta} \, \delta Q^2+\frac{\Theta}{\beta} \,\delta Q = 0.
\end{align}
We recover the differential equation of the harmonic LC oscillator in the linear limit with small signal amplitudes, which corresponds to $\delta Q^2 \rightarrow 0$.
Equation~\eqref{eq:DEQcompact} is known as the KdV equation~\cite{doi:10.1080/147864495086207391,Blyuss2002ChaoticBO1,DeaneNonlinearDO1,phdthesis1}. Its periodic solutions are given by cnoidal waves,
\begin{align}
\delta Q\,=\, \eta(m)+A_Q(m)\, \text{cn}^2 \bigl( \mu(m) \, t \,\mid m \bigr).
\end{align}
The parameter $\eta$ corresponds to the elevation of the wave \ie~the minimum of the signal relative to zero signal height. Due to the voltage offset, the total charge $Q(t)$ is shifted by the offset charge $Q_\text{offset}$. $A_Q$ is the absolute amplitude and $\mu$ the elliptic angular frequency. Here, $m$ is the elliptic parameter ($m \in [0,1]$), related to the eccentricity or 'elliptic modulus' $k$ according to $m=\sqrt{k}$.
By inserting the solution into \eqref{eq:DEQcompact}, we find the dependency of the quantities $\eta$, $A$, and $\mu$ on the eccentricity $m$ to be
\begin{subequations}
\begin{alignat}{4}
\eta(m)\, &= \,- \Theta + \dfrac{\Theta}{\sqrt{m^2-m+1}} \, \biggl( 1- 2\,m\biggr), \\
\label{eta}
A_Q(m) \,&=\, \dfrac{3\,m\, \Theta}{\sqrt{m^2-m+1}}, \\
\mu(m) \,&=\, \sqrt{\dfrac{\Theta}{\beta}} \biggl( 16 \bigl(m^2-m+1 \bigr) \biggr)^{-\frac{1}{4}},\\
T(m)\,&=\, \dfrac{2\, K(m)}{\mu(m)},
\label{T}
\end{alignat}
\end{subequations}
where $T$ is the period of the solution.
For a specific circuit system, $\mathcal{C}$, $\phi$, $V_{\text{offset}}$, $\nu$, and $L$ are determined and define the parameters $\Theta$ and $\beta$.
The elevation $\eta(m)$ connects to the linear limit at $m=0$ with $\eta(0)=0$, as the sinusoidal wave of period $T(0)=2\pi \sqrt{L\, C_0}=1/f_0$ with $C_0 = C(V_\text{offset})$ will be centered around zero. It is negative in the non-linear regime for $m>0$ as the solution corresponds to the AC contribution to the charge.

To find the solution for the AC voltage $u_C(t)$ across the capacitor, we can use equation~\eqref{U(Q(T))} and insert the solution for $\delta Q(t)$. In summary, the voltage signal of an unperturbed LC resonator with non-linear capacitance, modeled by equation \eqref{c(v)} is given by
\begin{align}
u_C(t) = \dfrac{1}{4\, \mathcal{C}^2 \,\phi} \, \big[ & \eta \, (\eta + 2 \Theta)  + 2 A_Q\, (\eta + \Theta) \, \text{cn}^2\bigr(\mu t  \, |\, m\bigr) \nonumber  \\
&+ A_Q^2 \, \text{cn}^4\bigr(\mu t \, | \, m\bigr)  \big].
\label{UC}
\end{align}
The parameters $\eta$, $A_Q$ and $\mu$ are plotted in Fig.~\ref{Parameters} with respect to the eccentricity $m$ using the fit obtained in Fig.~\ref{fig:singleresonator}.
Furthermore, \mbox{$L=(L_1^{-1}+L_2^{-1})^{-1}$} with $L_1 = \SI{338.5}{\micro \henry}$ and $L_2 = \SI{180.0}{\micro \henry}$ as experimentally implemented in the nSSH circuit.
Given the relation between voltage $u(t)$ and charge $\delta Q(t)$, we note that both oscillate with the same period $T$ and therefore have the same frequency $f=1/T$.
The total amplitude $u_{\text{pp}}$ and elevation $\eta_u$ of the voltage signal $u_C(t)$ are computed by evaluating maxima and minima of the solution in equation~\eqref{UC}.

%-------------------------------------------------------------------------------
% Figure 8
%-------------------------------------------------------------------------------
\begin{figure}[]
\centering
\includegraphics[width=1\linewidth]{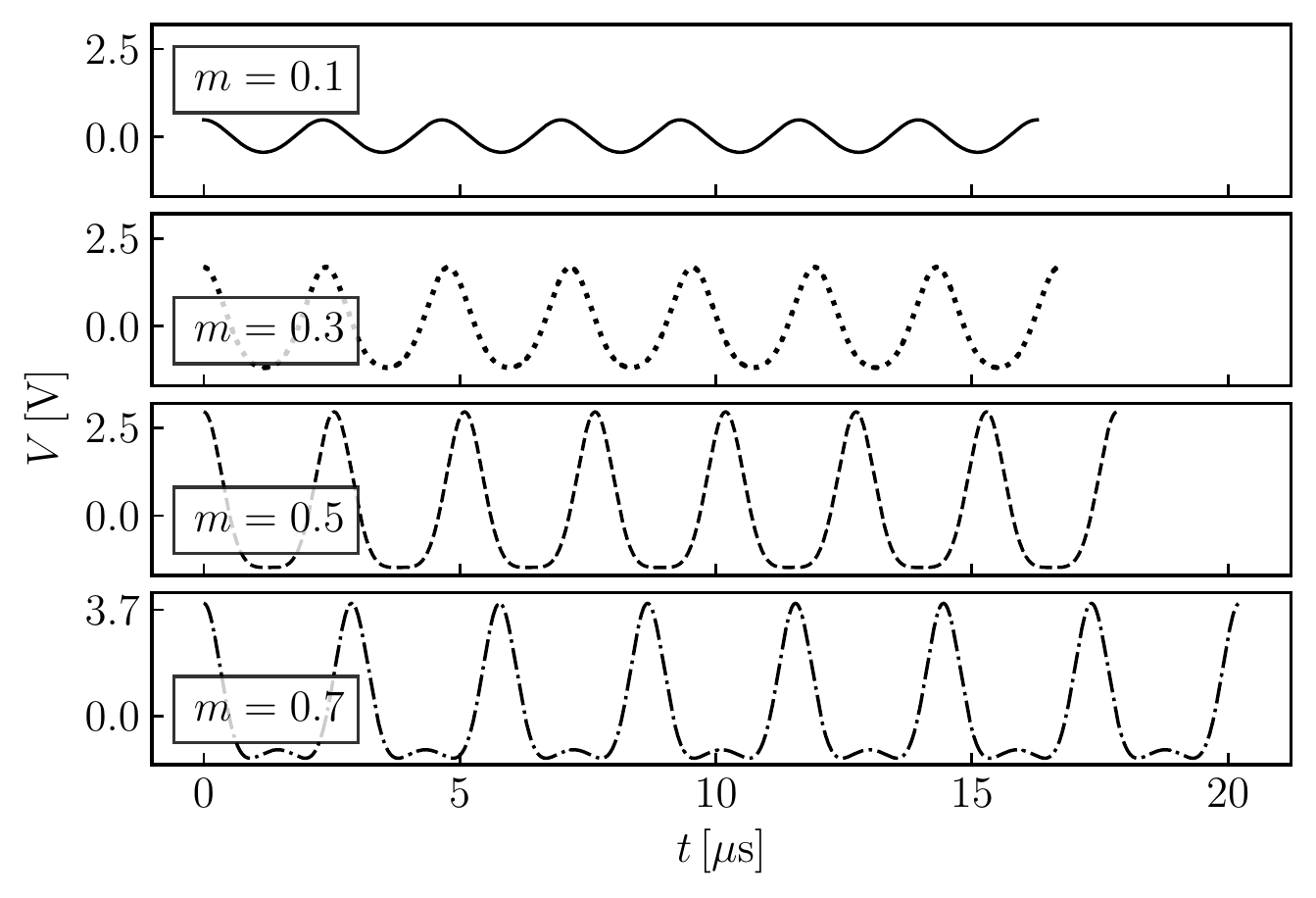}
\caption{Derived solution for the voltage signal of a single LC resonator \eqref{UC} for different eccentricity parameters $m$. For small $m$ \ie~the linear limit, the sinusoidal limit is recovered. For high $m$ the solution differs from the cnoidal wave solution as seen in the last plot.
}
\label{PlotSol}
\end{figure}

In Fig.~\ref{Parameters}~(a) we show the resonance frequency $f=1/T(m)$ of the solution of the single resonator, which depends on the eccentricity.
Each LC circuit in the single resonator approximation corresponding to one node of the full circuit network obtains a different resonance frequency, depending on its signal amplitude. At small $m < 0.3$ corresponding to voltage amplitdues $V<\SI{3}{\volt}$, we note that the resonance frequency stays approximately constant. We use this approximation to reconnect the single resonator formalism to the driving and assume that the single resonator frequency matches the driving frequency, which is unique throughout the whole circuit network. Within this approximation, the single resonator solution is applied as a solution of the EOM of the full circuit network including the external driving and describes the measured LCn state. Apart from the quality of the fit of equation~\eqref{c(v)}, this imposes the main limitation of the theoretical description of the LCn state. We denote this limit as the small amplitude limit, referring to the validity of the standard cnoidal wave solution.

Along with the resonance frequency, the whole spectrum shifts towards lower frequencies. This roots in a non-linear dispersion relation $\omega(k,|A|)$ with an additional amplitude dependency, as first proposed by Stokes for hydrodynamic waves, and encoded in cnoidal wave theory. It can be shown, that the resonance frequency always resides in the middle of the band gap.
As a reference point, the dashed line in Fig.~\ref{Parameters}~(a) indicates the cutoff frequency of the lower frequency band in the linear dispersion relation. In the SSH model, the lower and upper band of the dispersion relation correspond to states in the bulk while the midgap state resides at the boundary. The bandgap is a measure of robustness of the boundary state against disorder as well as a measure of localization strength with an attenuation factor of $\Delta = L_2/L_1$ for each unit cell. Subsequent research directions could be the investigation of the large amplitude limit where the non-linear midgap frequency resides inside the lower dispersion branch of the linear model, \ie~to study if the base frequency component delocalizes and to investigate interaction effects between bulk and boundary states.

From this estimated limitation we can deduce the optimal properties of the experimental setup. The operation point for the measurement was set to $V_{\text{offset}} = 2.5\,$V as a compromise to obtain high capacitances while maintaining amplitudes up to the implied upper limit for $\Delta L_{\text{nom}}= L_{2,\text{nom}}/L_{1, \text{nom}} \approx 0.545$. $\Delta L_{\text{nom}}$ was chosen such that the exponential decay is fast enough to reach the steady state limit before the reflection from the other end of the line interferes, see Suppl.~D.

\paragraph*{The soliton limit.}
In Fig.~\ref{PlotSol}, $7$ periods of the derived solution given by equation \eqref{UC} are displayed for different values of $m$. In agreement with the derived $m$ dependence, the period increases for higher eccentricities. In analog to the cnoidal wave solutions, the signal gets deformed resulting in a train of soliton like pulses. Though, for the single resonator approximation, the soliton limit for $m=1$ is not recovered (see section A). This is due to the fact that there exist no pulse solitons at rest.

% ------------------------------------------------------------------------------
%  Appendix D
% ------------------------------------------------------------------------------

\section{\\Experimental circuit setup}

Printed circuit boards (PCBs) were designed to obtain a versatile platform hosting various circuit component configurations in order to probe different SSH- and transmission line systems. The nSSH configuration consists of surface mounted inductors, varicap diodes, and capacitors.

\paragraph*{Capacitance to ground.}
Four Siemens variable capacitance (varicap) diodes BB512 in parallel circuit configuration connecting the voltage nodes to ground are used to obtain a non-linear on-site capacitance. The typical diode capacitance of the BB512 ($C_T=470\,$pF at $V_R=1\,$V and $f=1\,$MHz) states the upper limit of commercially available varicaps. The requirement of high capacitance values is imposed by two experimental circumstances. First, the resonance \ie~gap frequency is inversely proportional to the capacitance value. Lower frequencies are more convenient to handle within the chosen circuit setup and the signal can be resolved with a higher resolution relative to one period. Second, the parasitic contributions of the measurement setup are minimized. Arbitrary wave form generator (Agilent 33220A), oscilloscopes (PicoScope 4000 Series by PICOTech, AC-mode), BNC cables, and MFIA impedance analyzer (Zürich Instruments) impose additional capacitances of the order of $10^{-10}\,$F. To recover an identical effective on-site capacitance, all nodes are connected by $1\,$m BNC cables (RG58/U) to the oscilloscopes, inducing an additional capacitance of about $\approx 115\,$pF. The input node (e.g. $1$A for the transient measurement) is furthermore connected to the function generator by a BNC cable ($30\,$cm). The additional capacitance at the input node is approximated by measuring the resonance frequency with and without the connected input and compensated by using an approx. $40\,$cm BNC cable as connection to the oscilloscope instead of a $1\,$m BNC cable.

The influence of the MFIA could not be compensated entirely, leading to a shift towards smaller resonance frequencies for the impedance measurement (cf. Fig.1~(d) in main text).

\paragraph*{Inductors.}
Shielded, surface mounted inductors connect the voltage nodes. The requirements that need to be matched are: Relatively high inductance values to obtain a sufficiently low resonance frequency. The unavoidable serial resistance of the inductors should be kept as small as possible to obtain a maximized pulse life time, the serial resistance scales with the inductance value. To prevent spurious inductive coupling between inductors, they need to be magnetically shielded and have to be relatively small in order to design PCBs with sufficient spacings while keeping the total dimensions of the circuit manageable.
Above mentioned considerations resulted in the following choice of components:
	Coilcraft MSS1278-334KLD nominal values $ L_{1,\text{nom}} = 330\; \mu\text{H}$ and $R_{\text{DC}} = 487\; \text{m}\Omega$.
	Coilcraft MSS1260-184KLD nominal values $ L_{2,\text{nom}} = 180\; \mu\text{H}$ and $R_{\text{DC}} = 510\; \text{m}\Omega$.\\
To preserve translational symmetry the scatter of the absolute values of the circuit elements needs to be smaller than typical tolerances of commercially available components. To this end all components were precharacterized by a BK Precision 894 LCR-meter, resulting in typical inductance values of $L_1 = 334\,...\,343\, \mu$H and $L_2 = 176\,...\,184\, \mu$H measured at $f=430\,\si{\kilo\hertz}$.

\paragraph*{Dimension of the circuit.}
The experimental setup consists of $25$ unit cells. This number is sufficient to observe a clear steady state signal before the reflection from the end of the line induces disturbances.

\end{document}